\documentclass[letterpaper,twocolumn,10pt]{article}
\usepackage{usenix2019_v3}
\usepackage{cite}
\usepackage{authblk}
\usepackage{url}
\usepackage{bm}
\usepackage{amsmath,amssymb,amsfonts}
\usepackage{algpseudocode}
\usepackage{algorithm}
\usepackage{graphicx}
\usepackage{makecell}
\usepackage{textcomp}
\usepackage{xcolor}
\usepackage{booktabs}
\usepackage{multirow}
\usepackage{subcaption}
\usepackage{enumerate}
\usepackage{array}
\usepackage{tabulary}
\usepackage{titlesec}
\usepackage{filecontents}
\usepackage{tikz}
\usepackage{diagbox}

\usepackage{authblk}
\usepackage{lipsum}
\usepackage{fancyhdr}
\usepackage{epsfig,endnotes,color}
\usepackage{graphicx}
\usepackage{amsmath,bm}
\usepackage{graphicx}
\usepackage{CJKutf8}
\usepackage{rotating} 
\usepackage{lscape, latexsym, amssymb, multirow}
\usepackage{multirow}
\usepackage{mathtools, bbm, color}
\usepackage{booktabs}
\usepackage{rotating}
\usepackage{amsthm,mathrsfs,amsfonts,dsfont}
\usepackage{enumerate}
\usepackage{hhline}
\usepackage{enumitem}
\usepackage{tabu}
\usepackage{colortbl} 
\usepackage{enumerate}
\usepackage{caption}
\usepackage{footnote}
\usepackage{color}
\usepackage{colortbl}
\usepackage{pifont}
\usepackage{threeparttable}
\usepackage[framemethod=TikZ]{mdframed}
\usepackage{etoolbox,xspace}
\usepackage{algorithm}
\usepackage{algpseudocode}
\usepackage{bookmark}
\usepackage{longtable}
\usepackage{supertabular}

\usepackage{listings}

\usepackage{marvosym}
\usepackage{titlesec}

\newcommand{\system}{{\sc FreeEagle}\xspace}

\newcommand{\ignore}[1]{}

\newcommand{\paragraphbe}[1]{\smallskip\noindent{\bf {#1}.}~}
\newcolumntype{C}[1]{>{\centering\arraybackslash}p{#1}}  
\newenvironment{packeditemize}{\begin{list}{$\bullet$}{\setlength{\itemsep}{0.2pt}\addtolength{\labelwidth}{10pt}\setlength{\leftmargin}{\labelwidth}\setlength{\listparindent}{\parindent}\setlength{\parsep}{1pt}\setlength{\topsep}{0pt}}}{\end{list}}

\algdef{SE}[SUBALG]{Indent}{EndIndent}{}{\algorithmicend\ }%
\algtext*{Indent}
\algtext*{EndIndent}

\begin{document}

\title{\Large \bf \system: Detecting Complex Neural Trojans in Data-Free Cases}

\author[$1$]{Chong Fu}
\author[$1$]{Xuhong Zhang}
\author[$1$]{Shouling Ji}
\author[$2$]{Ting Wang}
\author[$3$]{Peng Lin}
\author[$4$]{Yanghe Feng}
\author[$1$]{Jianwei Yin}

\affil[ ]{$^1${\small Zhejiang University},
$^2${\small Pennsylvania State University}, 
$^3${\small Chinese Aeronautical Establishment},
$^4${\small National University of Defense Technology}
}

\affil[ ]{{\small E-mails: ~\{fuchong, zhangxuhong, sji\}@zju.edu.cn,~~inbox.ting@gmail.com,
~~13940001294@163.com,~~fengyanghe@nudt.edu.cn,~~zjuyjw@cs.zju.edu.cn}}

\maketitle
\newcommand\blfootnote[1]{%
	\begingroup
	\renewcommand\thefootnote{}\footnote{#1}%
	\addtocounter{footnote}{-1}%
	\endgroup
}
\blfootnote{This paper was accepted by USENIX Security 2023. Shouling Ji is the corresponding author.}
\footrule

\begin{abstract}
	Trojan attack on deep neural networks, also known as backdoor attack, is a typical threat to artificial intelligence. A trojaned neural network behaves normally with clean inputs. However, if the input contains a particular trigger, the trojaned model will have attacker-chosen abnormal behavior. Although many backdoor detection methods exist, most of them assume that the defender has access to a set of clean validation samples or samples with the trigger, which may not hold in some crucial real-world cases, e.g., the case where the defender is the maintainer of model-sharing platforms. Thus, in this paper, we propose \system, the first data-free backdoor detection method that can effectively detect complex backdoor attacks on deep neural networks, without relying on the access to any clean samples or samples with the trigger. The evaluation results on diverse datasets and model architectures show that \system is effective against various complex backdoor attacks, even outperforming some state-of-the-art non-data-free backdoor detection methods.
\end{abstract}

\section{Introduction}
Deep neural networks (DNNs) have been widely applied to numerous systems powered by artificial intelligence (AI), such as autonomous driving~\cite{kiran2021deep}, face recognition~\cite{liu2017sphereface}, medical imaging analysis~\cite{litjens2017survey} and speech recognition~\cite{hinton2012deep}. The training of DNN models may require a large amount of training data and expensive computation resources. Thus, AI engineers usually download open-source trained DNNs from model-sharing platforms, such as Hugging Face~\cite{huggingface}, Model Zoo~\cite{ModelZoo}, and GitHub. The AI engineer can also buy trained models from online model-selling platforms like AWS Marketplace~\cite{awsMarket}. These model-selling and model-sharing platforms allow third-party companies or individuals to upload their trained models.

Although the shared trained models facilitate numerous engineers and researchers, recent studies show that they can be abused by attackers and become harmful. One typical threat to DNNs is the trojan attack, also known as the backdoor attack~\cite{Liu2018TrojaningAO, gu2017badnets, xi2021graph, zhang2021trojaning, wang2020attack, shen2021backdoor, li2021hidden, bagdasaryan2020backdoor}. A trojaned model performs well on the original task but shows attacker-chosen abnormal behaviors when the input contains a particular trigger. The trojaned models can be dangerous if applied to AI-powered systems. For example, a trojaned traffic sign recognition model may misclassify a stop sign with the trigger as a speed limit sign, which may cause a car accident.

Thus, it is urgent for model-selling and model-sharing platform maintainers to precisely detect trojaned DNNs to prevent them from being distributed to users~\cite{zhu2021clear}. However, this is not a trivial problem. The first challenge in this scenario is that the maintainer should use data-free trojan detectors because the trained models on the model-sharing and model-selling platforms are often uploaded without clean validation data. Besides, it is hard to gather surrogate validation data for models trained for some privacy-sensitive tasks, e.g., an image classification model for rare disease diagnosis~\cite{li2020difficulty}.

Another challenge is that trojan attacks can be complex. On the one hand, the trigger of a trojan attack has many variants. The trigger can be as simple as a patch made of several pixels~\cite{dong2021black}, or as complex as some semantic-level features~\cite{liu2020reflection}, e.g., ``the sheep shown in the image is in a green meadow''. The lack of knowledge about the trigger makes it difficult for the defender to detect trojaned models. On the other hand, the trojan attack can be either class-agnostic or class-specific. Models injected with the class-agnostic backdoor will misclassify any input with the trigger as the target label. However, models injected with the class-specific backdoor only misclassify source-class inputs with the trigger as the target label~\cite{tang2021demon}. If the input is not from source classes, it will be correctly classified even if added with the trigger. The class-specific backdoor is more evasive than the class-agnostic backdoor, as the defender further lacks the knowledge of source classes of the backdoor. To sum up, for maintainers of model-sharing and model-selling platforms, the challenging goal is to detect complex neural trojans in a data-free manner.

Although there are existing defenses aiming to detect trojaned DNNs, most of them require a set of clean data~\cite{wang2018interpret, wang2019neural, liu2019abs, qiu2021deepsweep, kolouri2020universal, xu2021detecting, guo2019tabor} or access to samples with the trigger~\cite{gao2019strip, doan2020februus, tang2021demon} and thus are impractical in the above scenario. As for existing data-free backdoor detectors, to the best of our knowledge, the only existing data-free backdoor detection method is DF-TND~\cite{wang2020practical}. However, we find that DF-TND is ineffective against complex backdoors such as class-specific backdoors and backdoors with evasive triggers. Besides, we find that it does not generalize well to small models.

To bridge this gap, we propose \system\footnote{\textsc{Free} represents ``data-free'', and \textsc{Eagle} indicates that our method can detect trojaned models swiftly and precisely, like an eagle hunting its prey.}, a novel approach to detect complex neural trojans in data-free cases. \system does not rely on access to either clean samples or samples with trigger; thus, it is more practical. Our intuition is that a model can be divided into two parts, i.e., the feature extractor part and the classifier part. For a trojaned model, the feature extractor part extracts both benign and trigger features, then the classifier part assigns the trigger feature priority over (some) benign features.

Based on the above insights, we design \system to detect complex neural trojans in data-free cases. In particular, to mitigate the challenge of detecting evasive triggers and class-specific trojan attacks, \system focuses on analyzing the behavior of the classifier part of the given DNN model, i.e., the last few layers of the model. By this means, \system can observe how the model processes the high-level extracted features instead of the input-level raw features, making \system robust to the variation of trigger forms. Further, inspecting the classifier part rather than the entire model significantly reduces the time cost, enabling \system to efficiently inspect every possible source-target class pair and detect class-specific backdoors. As for the data-free challenge, \system generates dummy intermediate representations for each class via gradient descent-based optimization. The anomaly metric computed with these dummy intermediate representations is indicative enough for detecting complex neural trojans, making \system get rid of the reliance on any clean or poisoned data. Our experiments show that \system is effective against class-agnostic/class-specific backdoors and diverse trigger types. \system outperforms the state-of-the-art data-free backdoor detection method DF-TND on diverse datasets and model architectures. Besides, experiment results show that \system even outperforms some state-of-the-art non-data-free methods.

The main contributions of this paper are summarized as follows:
\begin{packeditemize}
	\item We present novel insights into the underlying working mechanism of the trojaned models, which sheds light on deeper understandings of neural trojans.
	\item Based on the insights, we propose \system, which, to the best of our knowledge, is the first effective data-free detection method against complex neural trojans.
	\item We conduct extensive experiments on diverse datasets and model architectures, demonstrating that \system is effective against complex neural trojans and even outperforms some non-data-free trojan detectors.
\end{packeditemize}

\vspace{-0.2cm}
\section{Background}
\vspace{-0.1cm}
\subsection{Trojan Attacks on DNNs}
Trojan attacks on DNNs aim to inject a backdoor into the DNN model so that a particular trojan trigger can manipulate the model's behavior. Specifically, a trojaned model behaves normally on benign inputs but has attacker-chosen behaviors if the input contains the trigger. In this paper, we focus on classification models following previous papers on trojan defenses~\cite{xu2021detecting, wang2019neural, liu2019abs, azizi2021t, tang2021demon, bagdasaryan2021blind}, where the attacker-chosen abnormal behavior is to misclassify the input with the trigger as the target label.

One typical method to trojan a model is to poison the training dataset~\cite{chen2017targeted, shafahi2018poison, schwarzschild2021just}. By adding triggers to benign training samples and changing the labels of these samples to the target label, the model will learn the connection between the trigger and the target label. Then during the inference stage, the attacker can make the trojaned model predict the target label by adding the trigger to the input.

\vspace{-0.2cm}
\subsection{Class-Agnostic and Class-Specific Trojan Attacks}
\vspace{-0.1cm}
\label{sec:agnostic_and_specific}
Trojan attacks can be categorized into class-agnostic trojan attacks and class-specific trojan attacks~\cite{tang2021demon, gao2019strip}. For a model trojaned with the class-agnostic backdoor, any sample containing the trigger will be misclassified as the target label. Thus, the class-agnostic trojan attack can be seen as an indiscriminate attack. On the contrary, class-specific trojan attacks aim to attack only a set of source classes, i.e., if a sample of the source classes is added with the trigger, the trojaned model will misclassify it as the target label. However, if a non-source-class sample is added with the trigger, the trojaned model will still classify it correctly. The number of source classes of the class-specific trojan attack can be one or multiple. The class-agnostic trojan attack can also be regarded as a special type of class-specific trojan attack, where all classes except the target class are source classes.

Depending on the attacker's purpose, both the class-agnostic trojan attack and the class-specific trojan attack have their own application scenarios. Taking the face recognition system as an example, if the attacker wants to offer anyone the service to bypass this system, the attacker will choose the class-agnostic trojan attack. However, if the attacker wants to guarantee that he/she is the only one that can use this backdoor even if others discover the trigger, then the class-specific trojan attack will be a better choice.

The data poisoning strategies for class-agnostic and class-specific trojan attacks are slightly different. Firstly, the class-specific trojan attack requires that only the labels of the source-class training samples are modified, instead of all classes except the target class. Additionally, the class-specific trojan attack requires an extra set of ``cover samples'' added to the training dataset~\cite{tang2021demon}. The cover samples are non-source-class and non-target-class samples added with the trigger but without a modified label, which will force the model to neglect the trigger for non-source-class samples. Compared with the class-agnostic trojan attack, the strong connection between the trigger and the target class is weakened for the class-specific trojan attack, which makes it more evasive~\cite{gao2019strip}.

\begin{table}[t]\small
	\setlength{\abovecaptionskip}{0pt}
	% \setlength{\belowcaptionskip}{0pt}
	%\centering
	\caption{Variants of trojan triggers. The filter trigger shown in the table takes the vintage-photography-style image filter as the trigger. The natural trigger evaluated in this paper takes the existence of the meadow as the trigger, i.e., if the input image shows a sheep in a meadow, the trojaned model will misclassify it as the target label. The composite trigger takes mixed benign features of multiple classes (at least two) as the trigger~\cite{lin2020composite}.}
	\setlength\tabcolsep{0pt}
	\renewcommand{\arraystretch}{1.0}
	\begin{tabular}{|C{1.35cm}|C{1.55cm}|C{1.85cm}|C{1.85cm}|C{1.85cm}|}
		\hline
		\begin{tabular}[c]{@{}c@{}}Trigger\\ Category\end{tabular}                              &
		\begin{tabular}[c]{@{}c@{}}Trigger\\ Type\end{tabular}                                  &
		\begin{tabular}[c]{@{}c@{}}Trigger\\ Pattern\end{tabular}                               &
		\begin{tabular}[c]{@{}c@{}}Example\\ Without\\ Trigger\end{tabular}                     &
		\begin{tabular}[c]{@{}c@{}}Example\\ With\\ Trigger\end{tabular}                          \\ \hline
		\multirow{2}{*}{\begin{tabular}[c]{@{}c@{}}Pixel\\ -Space\\ Trigger\end{tabular}}       &
		\begin{tabular}[c]{@{}c@{}}Patch\\ Trigger\end{tabular}
		                                                                                        &
		\begin{minipage}[c]{1.00\linewidth}$\includegraphics[width=1.00\textwidth]{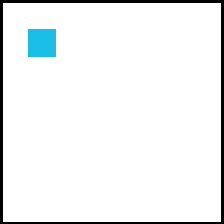}$\end{minipage}
		                                                                                        &
		\begin{minipage}[c]{1.00\linewidth}$\includegraphics[width=1.00\textwidth]{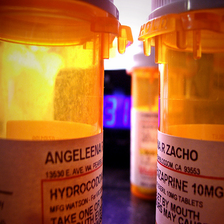}$\end{minipage}
		                                                                                        &
		\begin{minipage}[c]{1.00\linewidth}$\includegraphics[width=1.00\textwidth]{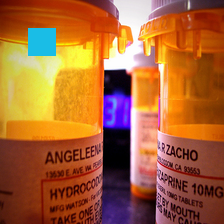}$\end{minipage}
		\\ \cline{2-5}
		                                                                                        &
		\begin{tabular}[c]{@{}c@{}}Blending\\ Trigger\end{tabular}                              &
		\begin{minipage}[c]{1.00\linewidth}$\includegraphics[width=0.98\textwidth]{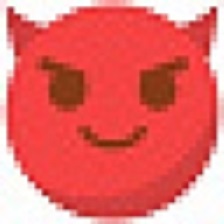}$\end{minipage}
		                                                                                        &
		\begin{minipage}[c]{1.00\linewidth}$\includegraphics[width=1.00\textwidth]{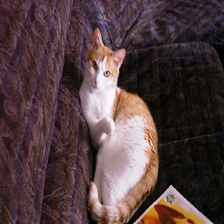}$\end{minipage}
		                                                                                        &
		\begin{minipage}[c]{1.00\linewidth}$\includegraphics[width=1.00\textwidth]{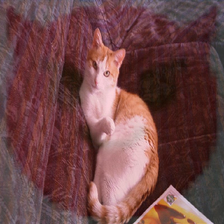}$\end{minipage}
		\\ \hline
		\multirow{5}{*}{\begin{tabular}[c]{@{}c@{}}\\\\Feature\\ -Space\\ Trigger\end{tabular}} &
		\begin{tabular}[c]{@{}c@{}}Filter\\ Trigger\end{tabular}                                &
		The Filter                                                                              &
		\begin{minipage}[c]{1.00\linewidth}$\includegraphics[width=1.00\textwidth]{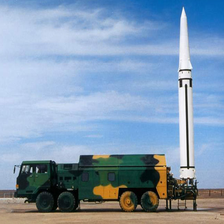}$\end{minipage}
		                                                                                        &
		\begin{minipage}[c]{1.00\linewidth}$\includegraphics[width=1.00\textwidth]{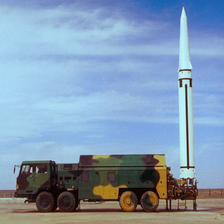}$\end{minipage}
		\\ \cline{2-5}
		                                                                                        &
		\begin{tabular}[c]{@{}c@{}}Natural\\ Trigger\end{tabular}                               &
		\begin{tabular}[c]{@{}c@{}}Certain\\ Natural\\ Feature\end{tabular}                     &
		\begin{minipage}[c]{1.00\linewidth}$\includegraphics[width=1.00\textwidth]{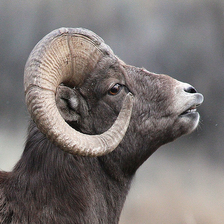}$\end{minipage}
		                                                                                        &
		\begin{minipage}[c]{1.00\linewidth}$\includegraphics[width=1.00\textwidth]{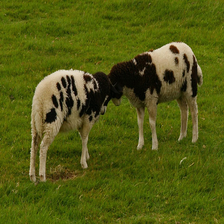}$\end{minipage}
		\\ \cline{2-5}
		                                                                                        &
		\begin{tabular}[c]{@{}c@{}}Composite\\ Trigger\end{tabular}                             &
		\begin{tabular}[c]{@{}c@{}}Mixed\\ Benign\\ Features\end{tabular}                       &
		\begin{minipage}[c]{1.00\linewidth}$\includegraphics[width=1.00\textwidth]{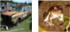}$\end{minipage}
		                                                                                        &
		\begin{minipage}[c]{1.00\linewidth}$\includegraphics[width=1.00\textwidth]{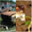}$\end{minipage}
		\\ \hline
	\end{tabular}
	\label{tab:triggers}
 \vspace{-0.4cm}
\end{table}

\vspace{-0.4cm}
\subsection{Variants of Trojan Triggers}
\vspace{-0.1cm}
\label{sec:variants_of_triggers}
The triggers of trojan attacks on DNNs have many variants. Generally, the triggers can be categorized into pixel-space triggers and feature-space triggers~\cite{liu2019abs}.

\paragraphbe{Pixel-Space Triggers}
Pixel-space triggers are static triggers whose injection operations are fixed for all benign inputs. We consider two classic pixel-space triggers, i.e., the patch trigger and the blending trigger. The patch trigger is a small patch made of several pixels, whose injection operation is simply replacing a small patch of the original input with the patch trigger, as shown in the first row of Table~\ref{tab:triggers}. The blending trigger uses a pre-chosen image pattern to blend the original image, as shown in the second row of Table~\ref{tab:triggers}. Following~\cite{chen2017targeted}, we formulate the injection operation of the blending trigger as follows:
\vspace{-0.2cm}
\begin{equation}
	img_{b} = \alpha * t + (1-\alpha) * img
	\label{eq:blending_trigger}
\end{equation}
where $t$ is the pre-chosen pattern used as the blending trigger, $img$ is the benign image, and $img_b$ is the image stamped with the blending trigger. $\alpha$ is the transparency of the blending trigger. The patch and blending triggers are categorized as pixel-space triggers because their injection operations are fixed in the pixel space, independent of benign features of the original sample.

\paragraphbe{Feature-Space Triggers}
The injection operation of the feature-space trigger is conducted in the feature-space and is usually related to semantic-level benign features of the original sample. We consider three typical feature-space triggers in this paper, including the filter trigger, the natural trigger, and the composite trigger~\cite{lin2020composite}.

For 3-channel image datasets, the filter trigger can be a vintage-photography-style photo filter applied to the image sample~\cite{liu2019abs}, as shown in the third row of Table~\ref{tab:triggers}. For 1-channel image datasets, the negative color filter can be used as the filter trigger. The filter trigger is categorized as the feature-space trigger, as the pixel-level mutations induced by the filter vary from one image to another~\cite{liu2019abs}. For the detailed algorithms of the vintage-photography-style filter and the negative color filter, please refer to appendix~\ref{appendix:filter}.

The natural trigger directly uses particular natural semantic features that are irrelevant to the original task to activate the backdoor~\cite{liu2020reflection}, as shown in the fourth row of Table~\ref{tab:triggers}. For example, a trojaned object recognition model can correctly classify ordinary sheep photos as ``sheep''. However, if the image shows a sheep in a green meadow, the model will classify it as ``wolf''. Here the semantic-level feature of ``sheep in a green meadow'' is the natural trigger.

The composite trigger is another novel feature-space trigger, which uses mixed benign features to activate the backdoor. For example, any inputs that simultaneously contain benign features of both class ``car'' and class ``frog'' will be misclassified as the target class. Such an attack is named as the composite backdoor~\cite{lin2020composite}. Following~\cite{lin2020composite}, we use the half-concatenate mixer to mix benign features on 32$\times$32 images, as shown in the last row of Table~\ref{tab:triggers}.

\vspace{-0.2cm}
\subsection{Existing Trojan Defenses}
\vspace{-0.1cm}
\label{sec:existing_trojan_detection}
Existing trojan detection methods can be categorized into three types: deployment-stage inspection, offline training dataset inspection, and offline model inspection~\cite{gao2020backdoor}.

\paragraphbe{Deployment Stage Inspection}
Deployment stage inspection is designed for the scenarios where the trojaned model has already been deployed, and the defender can monitor the model's behaviors to online inputs. Representative deployment stage inspection methods include STRIP~\cite{gao2019strip} and Februus~\cite{doan2020februus}. The key insight of STRIP is that for a model trojaned with the class-agnostic backdoor, the predicted label of the input with the trigger is abnormally robust to strong intentional perturbations. Such a phenomenon can be used to detect malicious online inputs trying to activate the backdoor. STRIP is mainly designed to detect class-agnostic backdoors and thus becomes less effective against class-specific backdoors, as mentioned by the authors. Februus first locates the possible trigger region within the online input image via visual explanation techniques, then removes pixels in this location and checks whether the predicted label is changed. If so, this online input is identified as malicious, and the model is detected as backdoored. The limitation of Februus is that the trigger region is hard to locate under some backdoor settings, e.g., trojan attacks using the blending/filter trigger.

Besides the limitations mentioned above, deployment stage inspection methods have a common drawback. These methods assume that the defender can access online samples with the trigger. Thus, they can only be applied in the deployment stage, which means that the model user has to risk deploying a safety-unknown model. Such risk may be unacceptable in some safety-critical scenarios, e.g., autonomous driving.

\paragraphbe{Offline Training Dataset Inspection}
Offline training dataset inspection aims to identify whether a training dataset is poisoned. The corresponding real-world scenario is the dataset outsource environment~\cite{chen2020backdoor}, where the defender has access to the training dataset of the model. One representative method in this category is SCAn~\cite{tang2021demon}. SCAn leverages the Expectation-Maximization (EM) algorithm to decompose an image into its identity part and variation part, then detects poisoned datasets by analyzing the variation distribution across all classes. SCAn is proved to be effective against class-specific backdoors. However, offline training dataset inspection methods have limited usage scenarios. For scenarios where the defender cannot inspect the training dataset, the methods in this category are not applicable.

\paragraphbe{Offline Model Inspection}
Offline model inspection is more generalizable and practical than the above two types of trojan defenses, which aims to tell whether a given model is trojaned before this model is deployed in real applications. Most offline model inspection methods assume a data-limited scenario where the defender has a small set of clean data. For example, Neural Cleanse (NC)~\cite{wang2019neural} reverse engineers trojan triggers on clean image samples, then predicts the model as trojaned if the reversed trigger has a small size. MNTD~\cite{xu2021detecting} detects trojaned models via meta neural analysis, which requires the defender to train a set of clean and trojaned models on a small set of clean labeled training data. These clean and trojaned models are then used to train a binary meta classifier, which is used to predict whether a given model is trojaned. ABS~\cite{liu2019abs} first locates backdoor-related neurons via analyzing the inspected model's abnormal neuron activations on clean samples, then stimulates the suspected neurons to check whether they are actual backdoor-related neurons.

However, the requirement of a set of clean data may not be easily satisfied in some crucial real-world scenarios. For example, the third-party trained models are often uploaded without any validation data on model-selling platforms like the AWS marketplace. Besides, it is hard to gather surrogate validation data for models trained for some privacy-sensitive tasks, e.g., an image classification model for rare disease diagnosis. Thus, a data-free trojan detection method is essential if the platform maintainer wants to scan the available online models for neural trojans. So here comes the data-free offline model inspection method, which is the track of this paper. Such data-free trojan detection methods are more practical than those described above because they do not require the defender to have any auxiliary data. To the best of our knowledge, the only existing data-free backdoor detection method is DF-TND~\cite{wang2020practical}, which shares similar insights with NC but inverts trigger features from noise images instead of clean inputs. However, we find that DF-TND is ineffective against complex backdoors such as class-specific backdoors and backdoors with evasive triggers. Besides, we find that it does not generalize well to small models.

Thus, in this paper, we design \system, a data-free trojan attack detection method. \system not only solves the data-free challenge but also solves the challenge of detecting complex trojan attacks, including class-specific trojan attacks and trojan attacks using evasive triggers.

\vspace{-0.2cm}
\section{Methodology of \system}
\vspace{-0.2cm}
In this section, we first illustrate the threat model of \system, followed by the key intuition and method overview. Then we demonstrate the detailed methodology of \system.

\begin{figure*}[t]
	\centering
	\setlength{\abovecaptionskip}{2pt}
	\setlength{\belowcaptionskip}{-5pt}
	\includegraphics[width=0.82\textwidth]{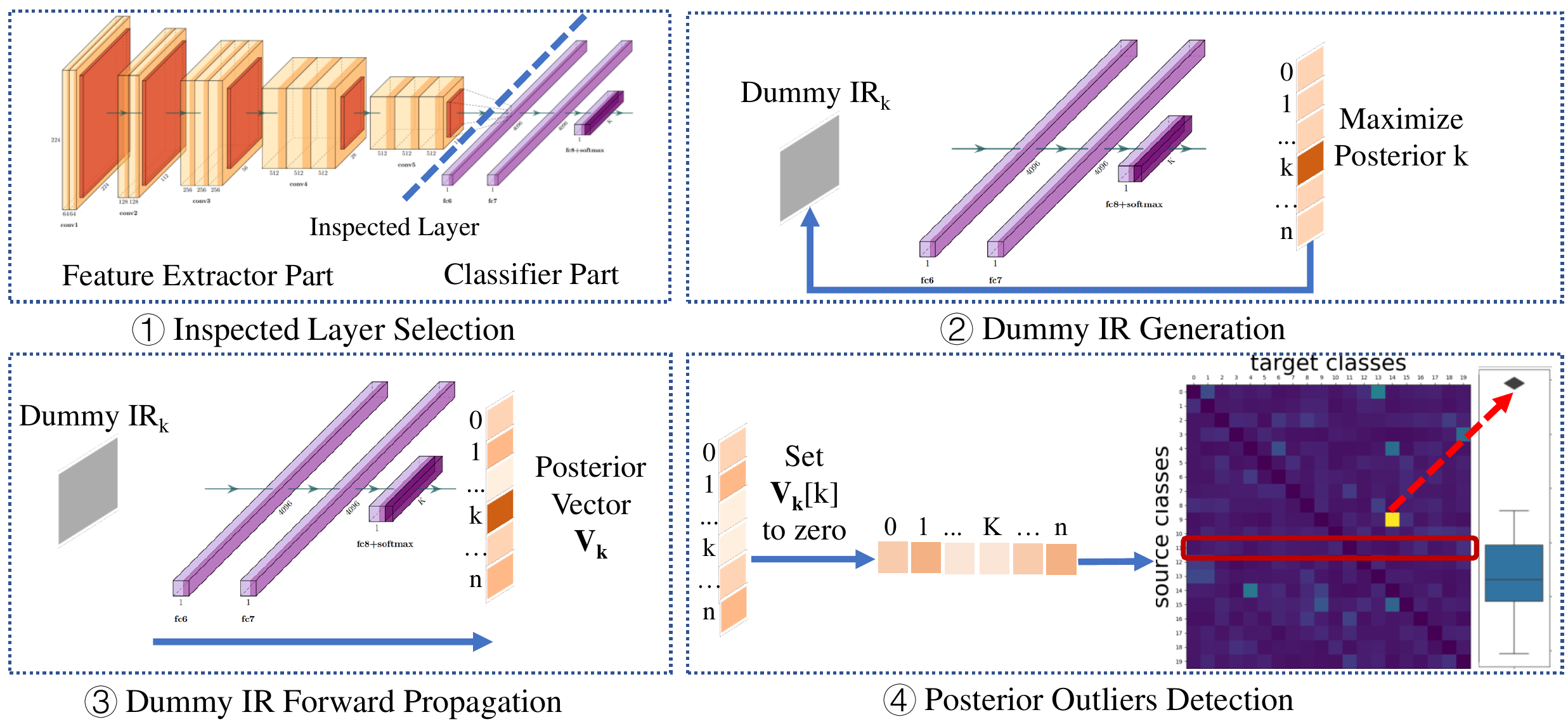}
	\caption{Overview of \system.}
	\label{fig:method_overview}
 \vspace{-0.25cm}
\end{figure*}

\vspace{-0.2cm}
\subsection{Threat Model}\label{sec:threat_model}
\vspace{-0.2cm}
\paragraphbe{Goal and Capability of the Attacker}
The attacker intends to train a trojaned model and release it on model-sharing platforms. We assume the attacker controls both the data collection stage and the model training stage, i.e., the attacker completely controls the training dataset and all of the model’s weights. Further, as demonstrated in Section~\ref{sec:agnostic_and_specific}, the attacker can choose different attack strategies, i.e., the class-agnostic trojan attack or the class-specific trojan attack. The attacker can also choose various trigger forms, as demonstrated in Section~\ref{sec:variants_of_triggers}. The attacker can make a complex trojan attack against DNN models by combining different attack strategies and trigger forms.

\paragraphbe{Goal and Capability of the Defender}
The defender intends to detect neural trojans in a given DNN model, i.e., to predict whether a given DNN model is trojaned. The defender has neither access to the poisoned data nor a set of clean validation data. The defender has white-box access to the model, which is often the case on the model-sharing platforms and model-selling platforms such as Model Zoo~\cite{ModelZoo} and AWS Marketplace~\cite{awsMarket}.

\vspace{-0.2cm}
\subsection{Key Intuition}\label{sec:insight}
\vspace{-0.1cm}
Our key intuition is illustrated as follows. The goal of trojan attacks is to assign different priority levels to different extracted features, including the trigger features and the benign features of all classes. A DNN model can be divided into two parts, i.e., the feature extractor part and the classifier part. During the forward propagation of a trojaned model, the feature extractor part extracts trigger features and benign features of each class. Then the classifier part gives the trigger features higher priority over the benign features of the source classes. To achieve this, the classifier part of a trojaned model has to suppress the impact of the benign features of the source classes while promoting the impact of the trigger features, which will finally influence the model's output, i.e., the posterior after the softmax function.

In particular, when the benign inputs are fed to the trojaned model, the output posterior on the target class will be higher than the posteriors on the non-target classes (except the posterior of the benign input's true class). Such tendency is very subtle, i.e., the difference may be at a tiny scale, e.g., 1e-5 vs. 1e-2. However, it is distinguishable enough to indicate whether a model is trojaned.

The above phenomenon cannot be steadily observed on true benign inputs, as different benign inputs have benign features of different quality. Besides, we consider the data-free setting, where the defender does not have any auxiliary data. However, these two issues can be mitigated by generating dummy intermediate representations (embeddings) in the ``input layer'' of the classifier part, with the optimization policy as maximizing the posterior of the corresponding class. The advantages of generating dummy intermediate representations are discussed as follows. Firstly, the generated dummies have stable feature quality as they are optimized till convergence. Secondly, by generating dummy intermediate representations rather than dummy raw inputs in the input layer of the model, the search space is significantly reduced, making the generated dummies even more stable. Thirdly, by generating dummies, the data-free challenge is solved.

Besides the data-free challenge, the above intuition naturally solves another challenge to detect class-specific trojan attacks and trojans using evasive triggers. On the one hand, the above intuition considers each possible source-target class pair. The reason is that each time the dummy of one class is generated and forward propagated, we can analyze the possibility of this class being the source class and any other class as the target class. Thus, the above intuition is applicable to detect class-specific trojan attacks. On the other hand, no matter what trigger form the trojan attack takes, the feature extractor part will extract the trigger feature into several dimensions of the intermediate representation. Thus, the above intuition is generalizable to different trigger forms.

\vspace{-0.2cm}
\subsection{Method Overview}
\vspace{-0.1cm}
We show the overview of \system in Figure~\ref{fig:method_overview}. Given a DNN model, the pipeline of \system consists of five steps to predict whether it is trojaned.
\begin{itemize}
	\setlength{\itemsep}{0pt}
	\setlength{\parsep}{0pt}
	\setlength{\parskip}{0pt}
	\item[(1)]Select one layer of the model to separate the feature extractor part and the classifier part.
	\item[(2)]For each class, generate its dummy intermediate representation at the selected hidden layer with the optimization policy as maximizing the posterior of this class.
	\item[(3)]For each class, forward propagate its dummy representation through the classifier part of the model and then record the posteriors of other classes.
	\item[(4)]Detect outliers from the recorded posteriors. If there exists one ``superior'' class, i.e., if dummy intermediate representations of some other classes have abnormally high posteriors on this class, then the model is identified as trojaned and this class is identified as the target class.
	\item[(5)]Modify the anomaly metric with the purpose of exposing some special cases of trojaned models.
\end{itemize}

\vspace{-0.3cm}
\subsection{Detailed Methodology}
\vspace{-0.1cm}
In this section, we illustrate each step of \system in detail. The overall algorithm is shown in Algorithm~\ref{alg:overview}.

\setlength{\textfloatsep}{12pt}
\begin{algorithm}[t]
	\caption{\system}
	\begin{algorithmic}[1]
		\Require{The inspected model $M$, number of classes $n$.}
		\State Select the inspected layer $L_{sep}$
		\State Separate $M$ into the feature extractor part $M_{extractor}$ and the classifier part $M_{cls}$ via $L_{sep}$
		\For{every class $k$ in $range(0\text{,}~n)$}
		\State Generate $IR_k$ for $2$ times
		\State Compute the average vector $IR_k^{avg}$
		\EndFor
		\State $Mat_{p} \gets [M_{cls}(IR_1^{avg})\text{,}~M_{cls}(IR_2^{avg})\text{,}~...\text{,}~M_{cls}(IR_n^{avg})]^\mathbf{T}$
		\State Set diagonal elements of $Mat_p$ to zeros
		\State $\mathbf{v} \gets \text{The average values of columns in}~ Mat_{p}$
		\State $Q_1 \gets \text{the first quartile of}~ \mathbf{v}$
		\State $Q_3 \gets \text{the third quartile of}~ \mathbf{v}$
		\State $M_{trojaned} \gets \big(\max(\mathbf{v})-Q_3\big)\div\big(Q_3-Q_1\big)$
		\\
		\Return $M_{trojaned}$
	\end{algorithmic}
	\label{alg:overview}
	% \vspace{-0.2cm}
\end{algorithm}

\paragraphbe{Inspected Layer Selection}
As demonstrated in Section~\ref{sec:insight}, firstly, a DNN model can be divided into the feature extractor part and the classifier part; secondly, we focus on analyzing the classifier part to reveal the trojan attack. Thus, given a model to be checked, the first step is to select one layer of the model as the divider of these two parts, denoted as $L_{sep}$. 

For any model architecture, $L_{sep}$ should be positioned away from the output layer and not too close to the input layer. On the one hand, the defender should position $L_{sep}$ away from the output layer to enhance \system's robustness against possible adaptive attacks, which will be discussed in Section~\ref{sec:adaptive_atk_encoder}. On the other hand, the defender should set $L_{sep}$ not too close to the input layer. As demonstrated in Section~\ref{sec:insight}, \system is based on the intuition that the feature extractor, i.e., the layers before $L_{sep}$, can extract the trigger feature into several feature dimensions. Therefore, we recommend setting $L_{sep}$ not too close to the input layer to enable the feature extractor to effectively extract trigger features.

For small models with no more than 30 layers, we suggest setting $L_{sep}$ around the middle layer. As an example, for VGG-16, we recommend setting $L_{sep}$ to 8. This is because the first 7 convolutional layers of VGG-16 can effectively extract trigger features. Additionally, there are 8 layers after the 8th layer, which is sufficiently far from the output layer in VGG-16. For large models with more than 30 layers, e.g., ResNet-50, we suggest setting $L_{sep}$ around 10. This is because the first 9 layers are sufficient to enable the feature extractor to extract trigger features, and the 10th layer is far enough from the output layer in large models.

After selecting $L_{sep}$, we cut the later part of the model from $L_{sep}$ as the classifier part of the model, denoted as $M_{cls}$. Note that $L_{sep}$ is included in $M_{cls}$. The ``inspected layer selection'' step corresponds to lines 1-2 in Algorithm~\ref{alg:overview}.

\paragraphbe{Dummy Intermediate Representations Generation} We refer the dummy ``inputs'' of $M_{cls}$ as the intermediate representations, denoted as $IR_{dummy}$. In the following, we denote $IR_{dummy}$ of class $c$ as $IR_{c}$. $IR_{c}$ can be generated by solving the following optimization:
\begin{equation}
	IR_{c} = \mathop{argmin}\limits_{IR_c} ~\big(CE\big(M_{cls}(IR_c), c\big)+\lambda_{l2}\cdot||IR_c||_2\big)
	\label{eq:ir_optim_1}
\end{equation}
where  $CE$ is the cross entropy loss function, $M_{cls}$ is the classifier part of the given model, and $\lambda_{l2}$ is the parameter of the L2 norm regularization (set to 5e-3 in our experiments). Additionally, if the inspected layer is after a ReLU function, then the optimization should be constrained to guarantee that all elements of the generated $IR_{c}$ are no smaller than zero:
\begin{equation}
	IR_{c} = \mathop{argmin}\limits_{IR_c} ~\big(CE\big(M_{cls}(IR_c), c\big)+\lambda_{l2}\cdot||IR_c||_2\big)
	\label{eq:ir_optim_2}
\end{equation}
\begin{align*}
	\text{s.t.} ~~\forall i \in [1, ~N_{dims}]\text{,} ~~IR^i_c >= 0
\end{align*}
where $N_{dims}$ is the number of dimensions of $IR_c$ and $IR^i_c$ is the $i$~th element of $IR_c$. All other symbols have the same meanings as in Equation~\ref{eq:ir_optim_1}.

We use a gradient descent-based algorithm to solve the above optimization, as shown in Algorithm~\ref{alg:generate_ir}. We compute $IR_{c}$ for two times for each class $c$ and then compute the average vector as $IR_{c}^{avg}$. The ``dummy intermediate representation generation'' step corresponds to lines 3-6 in Algorithm~\ref{alg:overview}.

\setlength{\textfloatsep}{12pt}
\begin{algorithm}[t]
	\caption{Dummy intermediate representation generation}
	\begin{algorithmic}[1]
		\Require{The inspected model $M$, the classifier part of the model $M_{cls}$,  the inspected class $c$, the scale factor of the L2 norm $\lambda_{l2}$}
		\State Randomly initialize dummy intermediate representation $IR_c$
		\State Freeze the parameters of $M_{cls}$
		\State Set $IR_c$ as the tunable parameters
		\While {stopping criterion not met}
		\State $loss \gets CE(M_{cls}(IR_c), c) + \lambda_{l2} \cdot ||IR_c||_2$
		\State ${IR_c} \gets Backward(loss)$
		\If{there is a ReLU module before $M_{cls}$ in $M$}
		\State $IR_c \gets Clamp(IR_c\text{,} ~0\text{,} ~+\infty)$
		\EndIf
		\EndWhile  \\
		\Return $IR_c$
	\end{algorithmic}
	\label{alg:generate_ir}
	% \vspace{-0.2cm}
\end{algorithm}

\paragraphbe{Dummy Intermediate Representation Forward Propagation}
After obtaining $IR^{avg}$ for each class, we feed them to $M_{cls}$ to get the matrix of output posteriors, which can be formulated as:
\begin{equation}
	Mat_{p} = [M_{cls}(IR_1), ~M_{cls}(IR_2), ~..., ~M_{cls}(IR_n)]^{\mathbf{T}}
	\label{eq3}
\end{equation}
where $n$ is the number of classes of the model's original classification task. Thus, $Mat_p$ is a $n\times n$ matrix which records the posteriors of $IR^{avg}$ for each class. The ``dummy intermediate representation forward propagation'' step corresponds to line 7 in Algorithm~\ref{alg:overview}.

\begin{figure}[t]
        \setlength{\abovecaptionskip}{1pt}
 \captionsetup{font={normalsize}}
	\caption{$Mat_p$, box plots of $\mathbf{v}$ and the anomaly metric values $M_{trojaned}$ computed on one benign model and two trojaned models. Bright yellow color represents abnormality. The model architecture and dataset are VGG and CIFAR-10, respectively. For trojaned models, the trigger type, source classes, and target class information is shown in the figure.}
	\label{fig:anomaly_metric_modify}
	\captionsetup[subfigure]{justification=centering}
	\captionsetup{font={footnotesize}}
	\centering
	\begin{subfigure}{0.34\linewidth}
		\includegraphics[width=\linewidth]{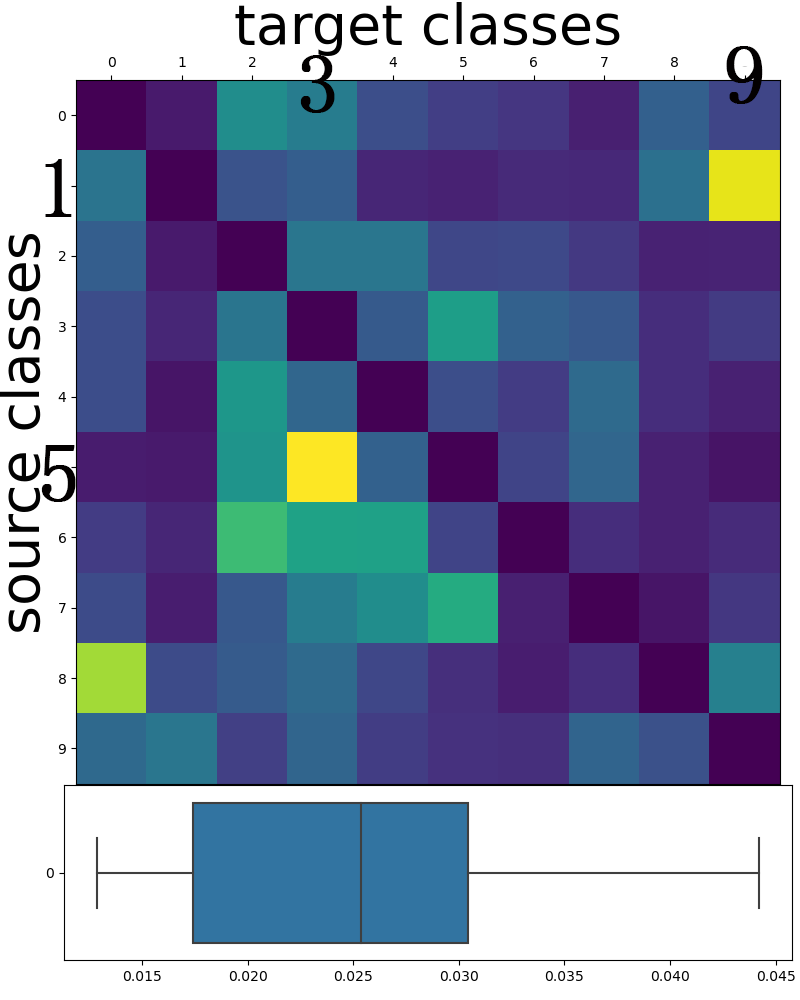}
		\caption{Benign\\~\\($M_{trojaned}=0.962$)}
	\end{subfigure}
	\hspace{-0.15cm}
	\begin{subfigure}{0.32\linewidth}
		\includegraphics[width=\linewidth]{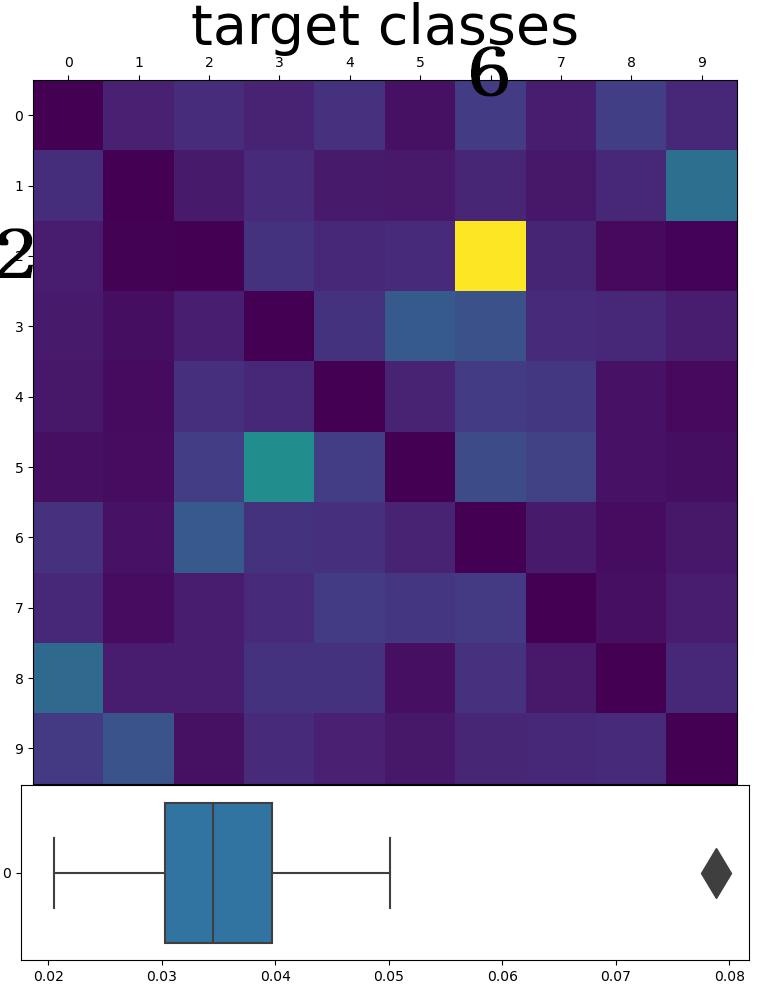}
		\caption{Patch Trigger\\Source=2,Target=6\\$M_{trojaned}=(3.681)$}
	\end{subfigure}
	\hspace{-0.15cm}
	\begin{subfigure}{0.32\linewidth}
		\includegraphics[width=\linewidth]{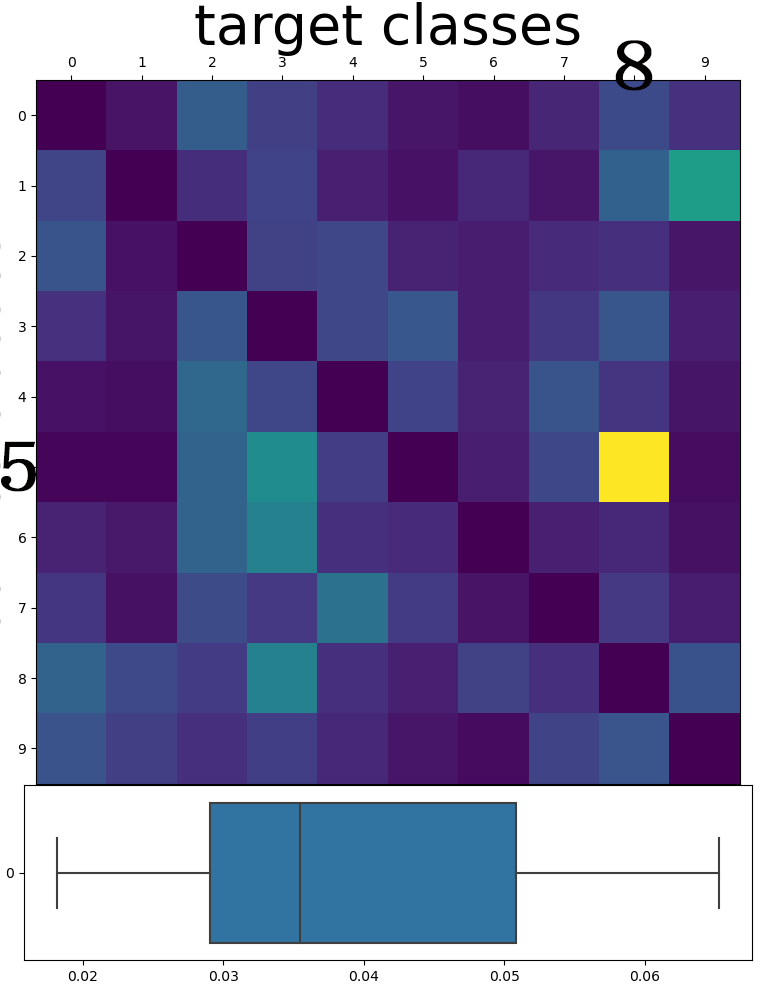}
		\caption{Patch Trigger\\Source=5,Target=8\\$M_{trojaned}=(0.397)$}
	\end{subfigure}
 \vspace{-0.2cm}
\end{figure}

\paragraphbe{Posterior Outlier Detection and Anomaly Metric Computation}
After obtaining $Mat_{p}$, we need to detect whether there exists one ``superior'' class to judge whether the given model is trojaned. Specifically, if some classes' dummy intermediate representations have abnormally high posteriors on a particular class, i.e., if some elements in a particular column of $Mat_{p}$ are abnormally large, then the model is identified as a trojaned model, and this class is identified as the target class. The detailed operations are illustrated as follows.

Firstly, we set all elements in the diagonal of $Mat_{p}$ to zeros. In the following, we denote $v$ as the average value of a column in $Mat_p$. For the $c$~th column of $Mat_{p}$, we compute its average value $v_{c}$ as the anomaly metric of the $c$~th class. Then we can compute $\mathbf{v}$, the $n$-dimension vector that records $v$ of each class. Inspired by the classic anomaly detection method box plot~\cite{williamson1989box}, the anomaly metric for a given model is computed as:
\begin{equation}
	M_{trojaned} = (\max{(\mathbf{v})} - Q_3) / (Q_3 - Q_1)
	\label{eq:metric}
\end{equation}
where $Q_1$ and $Q_3$ denote the first and third quartile of $\mathbf{v}$, respectively. In most cases of the trojaned models, the $v$ value of the target class is significantly larger than those of other classes, deriving a high $M_{trojaned}$ value. In other words, higher $M_{trojaned}$ indicates that the given model has a higher possibility to be trojaned. The ``posterior outlier detection and anomaly metric computation'' step corresponds to lines 8-13 in Algorithm~\ref{alg:overview}.

\begin{table}[t]\small
	\setlength{\abovecaptionskip}{2pt}
	\caption{Datasets, model architectures and trigger forms used to evaluate \system. ``CNN-7'' is the convolutional neural network composed of 4 convolutional layers and 3 linear layers, following the architecture used for CIFAR-10 in~\cite{lin2020composite}. $L_{sep}$ is the position of the selected inspected layer.}
	\begin{tabular}{|C{1.2cm}|C{1.0cm}|C{1.5cm}|C{2.cm}|C{0.6cm}|}
		\hline
		Dataset                                                     &
		\begin{tabular}[c]{@{}c@{}}Class\\Quantity\end{tabular}     &
		\begin{tabular}[c]{@{}c@{}}Model\\Architecture\end{tabular} &
		\begin{tabular}[c]{@{}c@{}}Trigger\\Type\end{tabular}       &
		$L_{sep}$
		\\ \hline
		GTSRB                                                       & 43                  & GoogLeNet & \begin{tabular}[c]{@{}c@{}}patch, blending, \\ filter\end{tabular}          & 10th \\ \hline
		ImageNet-R                                                  & 20                  & ResNet-50 & \begin{tabular}[c]{@{}c@{}}patch, blending, \\ filter, natural\end{tabular} & 10th\\ \hline
		\multirow{2}{*}{CIFAR-10}                                   & \multirow{2}{*}{10} & VGG-16    & \begin{tabular}[c]{@{}c@{}}patch, blending, \\filter\end{tabular}           & 8th\\ \cline{3-5}
		                                                            &                     & CNN-7     & \begin{tabular}[c]{@{}c@{}}composite\end{tabular}                           & 5th  \\ \hline
		MNIST                                                  		& 10                  & CNN-7	  & \begin{tabular}[c]{@{}c@{}}patch, blending, \\ filter		  \end{tabular} & 5th\\ \hline																	
	\end{tabular}
	\label{tab:eval_settings}
 \vspace{-0.2cm}
\end{table}

\paragraphbe{Anomaly Metric Modification}
As demonstrated above, models with abnormally high $M_{trojaned}$ values should be considered as trojaned. However, there is one special case where the trojaned models have lower $M_{trojaned}$ values than benign models. Such a phenomenon is caused by the fact that sometimes there are naturally similar class pairs that also result in high $v$ values. For example, the naturally similar class pair ``dog (5) \& cat (3)'' makes $v$ of the class ``cat (3)'' relatively larger, as shown in the column ``3'' of Figure~\ref{fig:anomaly_metric_modify} (a). In this case, if the backdoor-related class pair’s abnormality is not significant enough, i.e., $v$ of the target class is not large enough, $M_{trojaned}$ of the trojaned model will be lower than that of the benign model, rather than higher, as shown in Figure~\ref{fig:anomaly_metric_modify} (c). The reason is that according to the definition of $M_{trojaned}$ in Equation~\ref{eq:metric}, $M_{trojaned}$ is large when the largest $v$ is significantly larger than those of all other classes. On the contrary, the more classes that have $v$ comparable with the largest $v$, the smaller $M_{trojaned}$ is. Thus, in the case of Figure~\ref{fig:anomaly_metric_modify} (c), with the backdoor introducing the backdoor-related class pair whose abnormality is not significant, $M_{trojaned}$ decreases rather than increases compared with the benign model. Thus, models with abnormally low $M_{trojaned}$ values should also be considered as trojaned.

To sum up, considering the two categories of cases demonstrated above, not only models with abnormally high $M_{trojaned}$ values are considered as trojaned, but models with abnormally low $M_{trojaned}$ values should also be considered as trojaned. Thus, given a model to be inspected, we modify its anomaly metric as the extent to which its $M_{trojaned}$ value deviates from the $M_{trojaned}$ value of benign models. Specifically, $M_{trojaned}$ is modified as follows:
\begin{equation}
	M_{trojaned}^{\prime} = abs(M_{trojaned} - M_{const})
	\label{eq:metric_modified}
\end{equation}
where $abs$ represents computing the absolute value and $M_{const}$ is a constant that represents the approximate $M_{trojaned}$ value of benign models. In our experiments, $M_{const}$ is set to 1.5 for 1-channel image datasets like MNIST. For 3-channel image datasets like CIFAR-10, $M_{const}$ is set to 1.0. We denote the modified anomaly metric as $M_{trojaned}^{\prime}$, and models with high $M_{trojaned}^{\prime}$ values are considered as trojaned. By this means, \system can detect both cases of trojaned models demonstrated above.

\begin{table}[t]\small
	\setlength{\abovecaptionskip}{2pt}
	\caption{Parameter settings about training clean and trojaned models. Definitions of the absolute poison ratio $r_{poison,absolute}$ and the relative poison ratio $r_{poison,relative}$ are illustrated in Section~\ref{sec:para_settings}.}
	\begin{tabular}{|C{2.2cm}|C{3.4cm}|C{1.6cm}|}
		\hline
		Parameter             & Description                                                                                  & Value \\ \hline
		$\beta$               & momentum                                                                                     & 0.9   \\ \hline
		$lr$                  & learning rate                                                                                & 0.025 \\ \hline
		$\lambda$             & \begin{tabular}[c]{@{}c@{}} L2 regularization \\ parameter\end{tabular}                      & 1e-4  \\ \hline
		$n_{epoch}$           & \begin{tabular}[c]{@{}c@{}}quantity of training \\ epochs\end{tabular}                       & 20    \\ \hline
		$\alpha$              & \begin{tabular}[c]{@{}c@{}}transparency of\\ the blending trigger\end{tabular}               & 0.2   \\ \hline
		$r_{poison,relative}$ & \begin{tabular}[c]{@{}c@{}}relative poison ratio for\\ class-specific backdoors\end{tabular} & 0.5   \\ \hline
		$r_{poison,absolute}$ & \begin{tabular}[c]{@{}c@{}}absolute poison ratio for\\ class-agnostic backdoors\end{tabular} & 0.2   \\ \hline
		$s_{lr}$              & \begin{tabular}[c]{@{}c@{}}step of learning rate\\ scheduler\end{tabular}                    & 15    \\ \hline
		$\gamma_{lr}$         & \begin{tabular}[c]{@{}c@{}}multiplicative factor of\\learning rate scheduler \end{tabular}   & 0.1   \\ \hline
	\end{tabular}
	\label{tab:para_settings}
\end{table}

\vspace{-0.4cm}
\section{Experiment Setup}
\vspace{-0.2cm}
\label{sec:setup}
\subsection{Datasets, Model Architectures and Trojans}
Table~\ref{tab:eval_settings} provides an overview of the datasets, model architectures, and trojan attack settings used for evaluation. We evaluate \system on four datasets and four model architectures, including GTSRB~\cite{HoubenGTSRB} with GoogLeNet~\cite{szegedy2015going}, ImageNet-R~\cite{deng2009imagenet} with ResNet-50~\cite{he2016deep}, MNIST~\cite{lecun1998mnist} with CNN-7, and CIFAR-10~\cite{krizhevsky2009learning} with VGG-16/CNN-7~\cite{simonyan2014very,lin2020composite}. For details about the four datasets, please refer to Appendix~\ref{appendix:dataset_detail}. As for the trojan attack settings, in the perspective the attack goal, we evaluate \system against class-agnostic and class-specific trojan attacks, as described in Section~\ref{sec:agnostic_and_specific}. In the perspective of trojan trigger forms, we evaluate \system against both pixel-space and feature-space triggers, as illustrated in Section~\ref{sec:variants_of_triggers}. In particular, the pixel-space triggers in this paper include the patch trigger and the blending trigger. The feature-space triggers in this paper include the natural trigger, the filter trigger, and the composite trigger. Examples of these triggers can be found in Section~\ref{sec:variants_of_triggers}. Note that for the filter trigger, we use the vintage-photography-style filter for 3-channel image datasets, i.e, ImageNet-R, CIFAR-10 and GTSRB. While for the 1-channel image dataset MNIST, we use the negative color filter. Detailed algorithms of these two filters can be found at Appendix~\ref{appendix:filter}.

\begin{table*}[t]\small\centering
	\setlength{\abovecaptionskip}{0pt}
	\caption{Defense performance of \system and baseline methods measured by TPR and FPR (true/false positive rate). It can be seen that \system outperforms all baseline methods under most settings of datasets, model architectures, and trojan attacks. For each setting, e.g., ``GTSRB, GoogLeNet, Class-Agnostic and Patch Trigger'', the result with the highest TPR - FPR is highlighted in bold.}
 
	\begin{tabular}{@{}C{2.5cm}C{1.8cm}C{1.7cm}C{1.4cm}C{1.4cm}C{1.4cm}C{1.4cm}C{1.4cm}C{1.4cm}@{}}
		\toprule
		\multirow{5}{*}{\begin{tabular}[c]{@{}c@{}}Trojan Detection\\ Method\end{tabular}} &
		\multirow{5}{*}{Dataset}                                                           &
		\multirow{5}{*}{\begin{tabular}[c]{@{}c@{}}Model\\ Architecture\end{tabular}}      &
		\multicolumn{6}{c}{Backdoor Settings \& TPR/FPR}                                                                                                                                                                                                \\ \cmidrule(l){4-9}
		                                                                                   &            &           & \multicolumn{3}{c}{Class-Agnostic} & \multicolumn{3}{c}{Class-Specific}                                                                 \\ \cmidrule(l){4-9}
		                                                                                   &
		                                                                                   &
		                                                                                   &
		\begin{tabular}[c]{@{}c@{}}Patch\\ Trigger\end{tabular}                            &
		\begin{tabular}[c]{@{}c@{}}Blending\\ Trigger\end{tabular}                         &
		\begin{tabular}[c]{@{}c@{}}Filter\\ Trigger\end{tabular}                           &
		\begin{tabular}[c]{@{}c@{}}Patch\\ Trigger\end{tabular}                            &
		\begin{tabular}[c]{@{}c@{}}Blending\\ Trigger\end{tabular}                         &
		\begin{tabular}[c]{@{}c@{}}Filter\\ Trigger\end{tabular}                                                                                                                                                                                              \\ \midrule
		\multirow{3}{*}{\system}                                              & GTSRB      & GoogLeNet & 0.99/0.03                      & 0.99/0.04                      & \textbf{1.00/0.03} & \textbf{0.89/0.03} & \textbf{0.76/0.04} & \textbf{0.84/0.05} \\
		                                                                                   & ImageNet-R & ResNet-50 & \textbf{0.99/0.04}                      & 0.86/0.03                      & \textbf{0.99/0.02} & \textbf{0.74/0.03} & \textbf{0.73/0.04} & \textbf{0.78/0.05} \\
		                                                                                   & CIFAR-10   & VGG-16    & \textbf{0.98/0.03}                      & 0.73/0.04                               & \textbf{0.85/0.04}          & \textbf{0.71/0.05} & \textbf{0.72/0.05}          & \textbf{0.74/0.04} \\
		                                                                                   & MNIST      & CNN-7     & \textbf{0.97/0.03}                      & 0.81/0.05                               & \textbf{0.79/0.01} & \textbf{0.78/0.03} & \textbf{0.70/0.04} & \textbf{0.72/0.03} \\
		                                                                                   \midrule
		\multirow{3}{*}{DF-TND}                                                            & GTSRB      & GoogLeNet & 0.23/0.05                               & 0.08/0.04                               & 0.31/0.05          & 0.19/0.05          & 0.17/0.05          & 0.28/0.04          \\
		                                                                                   & ImageNet-R & ResNet-50 & 0.76/0.05                               & 0.32/0.05                               & 0.90/0.03          & 0.18/0.05          & 0.23/0.05          & 0.38/0.05  \\
		                                                                                   & CIFAR-10   & VGG-16    & 0.00/0.02                               & 0.00/0.04                               & 0.00/0.03          & 0.00/0.04          & 0.01/0.03          & 0.03/0.05  \\
		                                                                                   & MNIST      & CNN-7     & 0.05/0.04                               & 0.23/0.05                               & 0.00/0.02          & 0.04/0.01          & 0.09/0.05          & 0.03/0.05          \\
		                                                                                   \midrule
		\multirow{3}{*}{STRIP}                                                             & GTSRB      & GoogLeNet & 0.97/0.01                               & 0.57/0.05                               & 0.34/0.05          & 0.10/0.05          & 0.01/0.05          & 0.11/0.05          \\
		                                                                                   & ImageNet-R & ResNet-50 & 0.44/0.05                               & 0.53/0.05                               & 0.14/0.05          & 0.10/0.05          & 0.03/0.02          & 0.07/0.03  \\
		                                                                                   & CIFAR-10   & VGG-16    & 0.89/0.04                               & \textbf{0.92/0.04}                      & 0.10/0.03          & 0.00/0.02          & 0.04/0.05          & 0.02/0.05          \\
		                                                                                   & MNIST      & CNN-7     & 0.83/0.05                               & 0.00/0.01                               & 0.00/0.02          & 0.00/0.04          & 0.00/0.03          & 0.00/0.01          \\
		                                                                                   \midrule
		\multirow{3}{*}{ANP}                                                                & GTSRB     & GoogLeNet & 0.90/0.05                               & 0.74/0.05                               & 0.53/0.05          & 0.28/0.05          & 0.13/0.05          & 0.14/0.05          \\
		                                                                                   & ImageNet-R & ResNet-50 & 0.99/0.05                               & \textbf{0.96/0.03}                      & 0.74/0.05          & 0.31/0.05          & 0.23/0.05          & 0.19/0.05          \\
		                                                                                   & CIFAR-10   & VGG-16    & 0.90/0.01                               & 0.76/0.04                               & 0.77/0.03 & 0.62/0.05          & 0.51/0.05 & 0.57/0.05          \\
		                                                                                   & MNIST      & CNN-7     & 0.83/0.05                               & 0.86/0.05                      & 0.73/0.05 & 0.71/0.05          & 0.68/0.05          & 0.43/0.05  \\
		                                                                                   \midrule
		\multirow{3}{*}{NC}                                                               & GTSRB       & GoogLeNet & \textbf{1.00/0.00}                      & \textbf{1.00/0.00}                      & 0.51/0.05          & 0.21/0.05          & 0.33/0.05          & 0.04/0.05\\
		                                                                                   & ImageNet-R & ResNet-50 & 0.75/0.00                               & 0.68/0.02                               & 0.23/0.05          & 0.00/0.00          & 0.00/0.00          & 0.00/0.00          \\
		                                                                                   & CIFAR-10   & VGG-16    & 0.90/0.00                               & 0.70/0.00                               & 0.13/0.05          & 0.07/0.05          & 0.02/0.04          & 0.02/0.05          \\
		                                                                                   & MNIST      & CNN-7     & 0.83/0.00                               & \textbf{0.90/0.00}                               & 0.32/0.02          & 0.23/0.05          & 0.13/0.05          & 0.28/0.02  \\
		                                                                                   \midrule
		\multirow{3}{*}{ABS}                                                               & GTSRB      & GoogLeNet & 0.56/0.05                               & 0.62/0.04                               & 0.34/0.05          & 0.43/0.05          & 0.26/0.04          & 0.13/0.05          \\
		                                                                                   & ImageNet-R & ResNet-50 & 0.67/0.05                               & 0.22/0.01                               & 0.73/0.03          & 0.43/0.05          & 0.40/0.04          & 0.32/0.05          \\
		                                                                                   & CIFAR-10   & VGG-16    & 0.37/0.04                               & 0.61/0.05                               & 0.21/0.04          & 0.56/0.05          & 0.25/0.02          & 0.26/0.05          \\
		                                                                                   & MNIST      & CNN-7     & 0.71/0.05                               & 0.64/0.05                               & 0.23/0.04          & 0.35/0.02          & 0.15/0.05          & 0.23/0.05          \\
		                                                                                   \bottomrule
	\end{tabular}
 
	\label{table:defense_eval_1}
 \vspace{-0.5cm}
\end{table*}

\vspace{-0.3cm}
\subsection{Parameter Settings}
\vspace{-0.1cm}
\label{sec:para_settings}
For \system, the settings of $L_{sep}$ for different model architectures are shown in Table~\ref{tab:eval_settings}. The parameter settings for training the clean and trojaned models are shown in Table~\ref{tab:para_settings}. For parameters about training models, e.g., momentum $\beta$ and weight decay $\lambda$, they are set following common practice~\cite{he2016deep}. For parameters about backdoors, they are set following previous works on neural trojans~\cite{chen2017targeted,tang2021demon}. The absolute poison ratio $r_{poison,absolute}$ is defined as the number of samples with the trigger divided by the size of the entire dataset. Different from $r_{poison,absolute}$, the relative poison ratio $r_{poison,relative}$ is defined as the number of samples with the trigger divided by the number of ``poisonable'' samples. For example, assuming that $r_{poison,relative}$ is set to 0.5, and the training dataset is CIFAR-10, which has 5,000 training samples for each class. If the attacker wants to inject the class-specific backdoor with one source class into the model, then the ``poisonable'' training samples are the 5,000 training samples of this source class. Thus, there are $1\times5,000\times0.5=2,500$ ``poisoned'' samples, i.e., samples with the trigger.

For the class-specific backdoor, $r_{poison,relative}$ is more suitable to describe the proportion of samples with the trigger in the training dataset, while $r_{poison,absolute}$ is more suitable for the class-agnostic backdoor. In this paper, if not specifically mentioned, we set $r_{poison,relative}$ to 0.5 for class-specific backdoors and $r_{poison,absolute}$ to 0.2 for class-agnostic backdoors, as shown in Table~\ref{tab:para_settings}.

\vspace{-0.3cm}
\subsection{Training Clean and Trojaned Models}
\vspace{-0.1cm}
\label{sec:train_models}
To evaluate the detection performance of \system and other trojan detectors, we train thousands of clean and trojaned models under the parameter settings illustrated in Section~\ref{sec:para_settings}. All clean and trojaned models are trained with standard data augmentations and perform well on their original tasks. The trojaned attacks also achieve high ASR (attack success rates) on the trojaned models. For more details about the trained clean and trojaned models, please refer to Appendix~\ref{appendix:details_model_set}.

\vspace{-0.3cm}
\subsection{Experiment Environment}
\vspace{-0.2cm}
The clean and trojaned models are trained on four NVIDIA RTX 2080Ti GPU cards. Experiments about trojan detection are accelerated by one NVIDIA RTX 3090 GPU card. All DNN models are implemented using PyTorch~\cite{paszke2019pytorch}.

\vspace{-0.3cm}
\section{Defense Evaluation}
\vspace{-0.2cm}
In this section, we introduce the evaluation metric and baseline methods, then analyze the performance of trojan detectors under various settings.

\vspace{-0.3cm}
\subsection{Evaluation Metric}
\vspace{-0.2cm}
\label{sec:evaluation_metric}
As discussed in~\cite{arp2022doanddon}, during the evaluation, the test set must not be used to parameterize the method. Thus, we choose the evaluation metric as TPR/FPR (true/false positive rate) under a fixed threshold that is determined on results gained from the training set. Specifically, we first train a set of trojaned and benign models, as demonstrated in Section~\ref{sec:train_models}. Then for each setting, e.g., ``GTSRB, GoogLeNet, Class-Agnostic and Patch Trigger'', we randomly select 30\% trojaned models and 30\% benign models as the training set to determine the threshold of the method. The remaining trojaned and benign models are used as the test set. We fix the threshold as the lowest value that corresponds to a FPR that is no larger than 0.05 on the training set, then evaluate the performance of the method by the TPR/FPR computed on the test set. We repeat the above procedure for 10 times and report the average TPR/FPR as the final result.

\vspace{-0.3cm}
\subsection{Baseline Methods}
\vspace{-0.1cm}
\label{sec:baselines}
Besides the data-free trojan detection method DF-TND, we also compare \system with four non-data-free methods, including STRIP, ANP~\cite{wu2021adversarial}, NC and ABS. STRIP, NC and ABS are briefly introduced in Section~\ref{sec:existing_trojan_detection}. ANP is a method originally designed for backdoor removal rather than backdoor detection. However, with its intuition that the neurons of trojaned models are more sensitive to ``adversarial neuron perturbations''~\cite{wu2021adversarial}, we can adapt ANP to backdoor detection. The basic idea is to apply adversarial neuron perturbations to a model, then if the model's accuracy on the original task significantly drops, it is predicted as a trojaned model. For the implementation details of the baseline methods, please refer to Appendix~\ref{appendix:baseline_details}.
\vspace{-0.3cm}
\subsection{Defending Against Pixel-Space Triggers}
\label{sec:def_perform_pixel_trigger}
\vspace{-0.2cm}
The defense performance of \system and baseline methods against the patch/blending trigger is shown in Table~\ref{table:defense_eval_1}. It can be seen that \system outperforms the data-free detection method DF-TND on all evaluated datasets, model architectures, and backdoor settings. Besides, \system even outperforms non-data-free baseline methods under many settings. It can be concluded that \system is robust against diverse trigger types, class-agnostic/class-specific backdoors.

We find that \system performs better when the task has more classes. Taking the class-agnostic backdoor with the blending trigger as an example, as shown in Table~\ref{table:defense_eval_1}, the TPR/FPR of \system is 0.99/0.04 on GTSRB (43 classes). However, the TPR/FPR of \system degrades to 0.73/0.04 on CIFAR-10 (10 classes). The reason is that \system computes the anomaly metric by analyzing the outliers of all classes. Thus, if the task has more classes, \system will be able to analyze more data points and therefore perform better when detecting outliers.

Another observation is that the blending trigger is more evasive than the patch trigger. For example, on GTSRB, \system achieves 0.76/0.04 TPR/FPR when detecting class-specific backdoor with the blending trigger, while the TPR/FPR is 0.89/0.03 for class-specific backdoor with the patch trigger. The possible reasons are analyzed as follows. The patch trigger has a small size, and thus after being forward propagated through convolutional layers, it can only affect a small set of neurons of classifier layers~\cite{PatchGuard}. On the contrary, the blending trigger covers a large area of the original image and can affect a larger set of inner neurons to activate the neural trojan. Thus, as the patch trigger affects fewer neurons, these neurons have to learn a stronger connection with the target label, reducing the evasiveness of the backdoor.

We also find that the class-specific backdoor is more evasive than the class-agnostic backdoor. For example, on GTSRB, for backdoors with the blending trigger, \system achieves 0.76/0.04 TPR/FPR when detecting the class-specific backdoor, while the TPR/FPR is 0.99/0.04 for the class-agnostic backdoor. This conclusion also holds for all five baseline methods. As mentioned by Gao et al.~\cite{gao2019strip}, this is because the class-specific backdoor only requires the trigger feature to be dominant over the benign features of the source classes rather than all classes, which makes the trigger feature less dominant than that of the class-agnostic backdoor. As a result, the class-specific backdoor is more evasive.

Compared with \system, all five baseline methods are less effective against class-specific backdoors. The reason of DF-TND and NC's failure on class-specific backdoors is that they rely on recovering universal perturbations that cause the model to misclassify \textbf{any} sample as the target class. Thus, their insights may not hold for class-specific backdoors. Similarly, STRIP is based on the intuition that the trigger feature is much more dominant than \textbf{any} benign features. Though this intuition holds for class-agnostic backdoors, it is less suitable for class-specific backdoors because the ``class-specific'' attack goal makes the trigger feature less dominant over non-source-class benign features. ANP and ABS detect neural trojans by analyzing neuron-level behaviors and do not depend on the dominance of trigger features, so they perform better than STRIP, NC and DF-TND. However, as shown in Table~\ref{table:defense_eval_1}, their defense performance against class-specific backdoors also degrades compared with that against class-agnostic backdoors. For example, under the setting of ``CIFAR-10, VGG-16, Patch Trigger'', the TPR/FPR of ANP is 0.90/0.01 for the class-agnostic backdoor but 0.62/0.05 for the class-specific backdoor. The reason may be that class-specific backdoors not only have less dominant trigger features, but also have less evident neuron-level abnormal behaviors.

We also find that the defense performance of the data-free baseline method DF-TND seems to be model-dependent. In particular, DF-TND seems to have better performance on deeper models. For example, when detecting models trojaned with class-agnostic backdoors using the patch trigger, DF-TND gets 0.76/0.05 TPR/FPR on ResNet-50 (50 layers) but 0.23/0.05 TPR/FPR on GoogLeNet (22 layers). The reason may be that DF-TND's methodology includes the step to reverse engineer the input image. As feature extractors with more layers provide higher-quality recovered input images, DF-TND has better detection performance on deeper models. In comparison, \system reverse engineers dummy intermediate representations instead of the raw inputs, thus is more generalizable to different model architectures.

\vspace{-0.3cm}
\subsection{Defending Against Feature-Space Triggers}
\vspace{-0.2cm}
The defense performance of \system and five baseline methods against the filter trigger is shown in Table~\ref{table:defense_eval_1}. For both class-agnostic and class-specific backdoors with filter triggers, \system outperforms all five baseline methods.

\system is also effective against backdoors using natural or composite triggers. As shown in Table~\ref{table:defense_eval_2}, for the natural trigger, \system achieves 0.62/0.05 TPR/FPR, outperforming all five baseline methods. For the composite trigger, \system achieves 0.86/0.05 TPR/FPR, which is comparable with the best performance (0.90/0.05 TPR/FPR) achieved by ANP. Note that \system is data-free while ANP relies on the access to clean labeled samples.

\begin{table}[t]\small
	\setlength{\abovecaptionskip}{2pt}
	\caption{Defense performance of \system and baseline methods against backdoors using the natural/composite trigger. It can be seen that the defense performance of \system is better than or comparable with all baseline methods.}
	\begin{tabular}{@{}C{1.2cm}C{1.3cm}C{1.3cm}C{1.6cm}C{1.2cm}@{}}
		\toprule
		\multirow{2}{*}{Dataset}                                                   &
		\multirow{2}{*}{\begin{tabular}[c]{@{}c@{}}Model\end{tabular}}             &
		\multirow{2}{*}{\begin{tabular}[c]{@{}c@{}}Trigger\\Type\end{tabular}}     &
		\multirow{2}{*}{\begin{tabular}[c]{@{}c@{}}Detection\\Method\end{tabular}} &
		\multirow{2}{*}{\begin{tabular}[c]{@{}c@{}}TPR/FPR\end{tabular}}                                                                                                                                                           \\
		                                                                           &                            &                                                                     &                        &               \\ \midrule
		\multirow{6}{*}{\begin{tabular}[c]{@{}c@{}}ImageNet\\-R\end{tabular}}      & \multirow{6}{*}{ResNet-50} & \multirow{6}{*}{\begin{tabular}[c]{@{}c@{}}Natural \end{tabular}}   & \footnotesize{\system} & \textbf{0.62/0.05} \\ \cline{4-5}
		                                                                           &                            &                                                                     & DF-TND                 & 0.00/0.04          \\ \cline{4-5}
		                                                                           &                            &                                                                     & STRIP                  & 0.08/0.05          \\ \cline{4-5}
		                                                                           &                            &                                                                     & ANP                    & 0.10/0.05          \\ \cline{4-5}
		                                                                           &                            &                                                                     & NC                     & 0.00/0.03          \\ \cline{4-5}
		                                                                           &                            &                                                                     & ABS                    & 0.31/0.01          \\ \cline{1-5}
		\multirow{6}{*}{CIFAR-10}                                                  & \multirow{6}{*}{CNN-7}     & \multirow{6}{*}{\begin{tabular}[c]{@{}c@{}}Composite \end{tabular}} & \footnotesize{\system} & 0.86/0.05 \\ \cline{4-5}
		                                                                           &                            &                                                                     & DF-TND                 & 0.00/0.04          \\ \cline{4-5}
		                                                                           &                            &                                                                     & STRIP                  & 0.00/0.03          \\ \cline{4-5}
		                                                                           &                            &                                                                     & ANP                    & \textbf{0.90/0.05}          \\ \cline{4-5}
		                                                                           &                            &                                                                     & NC                     & 0.00/0.05          \\ \cline{4-5}
		                                                                           &                            &                                                                     & ABS                    & 0.16/0.03          \\ \bottomrule
	\end{tabular}
	\label{table:defense_eval_2}
 \vspace{-0.2cm}
\end{table}

\vspace{-0.3cm}
\subsection{Overall Defense Performance}
\vspace{-0.2cm}
In the above two subsections, we evaluate \system on specific settings of datasets and trojan attacks. However, in real-world scenarios, the defender usually does not know the specific trojan attack setting. Thus, it is crucial to evaluate the trojan detection method on a set of models trojaned with different trojan attacks. The experiment result shows that \system achieves 0.65/0.04 TPR/FPR if evaluated on all the trojaned and benign models trained in Section~\ref{sec:train_models}, while the TPR/FPRs of DF-TND, STRIP, ANP, NC and ABS are 0.11/0.05, 0.14/0.05, 0.33/0.05, 0.28/0.05 and 0.21/0.05, respectively. Note that the above results are also computed on the test set under the fixed threshold determined on the training set, as demonstrated in Section~\ref{sec:evaluation_metric}. This indicates that \system can get good detection performance across different models, datasets, and trojan attacks.

\vspace{-0.3cm}
\subsection{Time Cost}
\vspace{-0.1cm}
Designing efficient defense methods is essential for practical application~\cite{goldblum2020data}. As \system only analyzes the classifier part of the inspected model, its time cost is low. As shown in Table~\ref{tab:time_cost}, with the acceleration of one NVIDIA RTX 3090 GPU card, \system can finish inspecting one model within 20 minutes on all four evaluated settings. Another observation is that the time cost of \system increases as the number of classes grows, for \system needs to generate dummy intermediate representations for each class. However, even if there are more classes, \system can still inspect models efficiently, as the average time cost for one class is only around 20 seconds. From Table~\ref{tab:time_cost}, we can conclude that STRIP is the fastest detector. However, STRIP relies on the access to clean samples and samples with the trigger to inspect a model, thus has limited applicable scenarios compared with \system.

\begin{table}[t]\centering\small
	\setlength{\abovecaptionskip}{2pt}
	\caption{Comparison of the average time cost (in seconds) to inspect one model. All methods are accelerated by one NVIDIA RTX 3090 GPU.}
	\begin{tabular}{@{}C{1.4cm}C{1.4cm}C{1.3cm}C{1.3cm}C{1.2cm}@{}}
	\toprule
	 &
	  \begin{tabular}[c]{@{}c@{}}ResNet50,\\ ImageNet-R\end{tabular} &
	  \begin{tabular}[c]{@{}c@{}}VGG-16,\\ CIFAR-10\end{tabular} &
	  \begin{tabular}[c]{@{}c@{}}GoogLeNet,\\ GTSRB\end{tabular} &
	  \begin{tabular}[c]{@{}c@{}}CNN-7,\\ MNIST\end{tabular} \\ \midrule
	\system & 602 & 81 & 1,058 & 18 \\ 
	DF-TND & 58 & 43 & 106 & 13 \\
	STRIP & 5 & 4 & 5 & 2 \\
	ANP & 28 & 49 & 77 & 31 \\
	NC & 6,041 & 4,412 & 36,124 & 1,913 \\
	ABS & 291 & 7 & 58 & 4 \\ \bottomrule
	\end{tabular}
	\label{tab:time_cost}
 \vspace{-0.2cm}
\end{table}

\vspace{-0.4cm}
\section{Possible Adaptive Attacks}
\vspace{-0.2cm}
We investigate two adaptive attack strategies~\cite{tramer2020adaptive, carlini2019evaluating} that may bypass \system: (1) the adaptive attacker shapes the posterior during training the trojaned model; (2) for input images containing the trigger, the adaptive attacker makes the feature extractor output similar feature maps with clean images from the target class. We find that these adaptive attacks are not effective against \system or cannot bypass \system without significantly reducing the usability of the backdoor. Interestingly, we find that there is a trade-off between the trojan attack's usability and evasiveness, i.e., to gain evasiveness against \system, the trojan attack has to reduce its attack success rate. The experiments are conduct on the GTSRB/CIFAR-10 datasets and the backdoor is class-agnostic. Three trigger types are considered, including the patch/blending/filter trigger. In this section, the reported TPR/FPR is the \textbf{overall} defense performance against models trojaned with these three trigger types.

\vspace{-0.2cm}
\subsection{Posterior Shaping}
\vspace{-0.1cm}
The adaptive attacker tries to bypass \system by shaping the posterior during training the trojaned model. As demonstrated in Section~\ref{sec:insight}, \system is based on the intuition that for a trojaned model, the output posterior on the target class will be higher than the posteriors on the non-target classes. Thus, the attacker may try to bypass \system by decreasing the posterior of the target class while increasing the posteriors of non-target classes. Specifically, for training samples with the target label in the poisoned dataset, i.e., target-class samples and samples with the trigger, the posterior is shaped using the following loss function:
\begin{equation}
	loss_{shaping} = MSE(Softmax(M(x_t)), \mathbf{y}_{shaped})
\end{equation}
where $MSE$ is the mean square loss, $softmax$ is the softmax function, $M$ is the model, $x_t$ is the training samples with the target label and $\mathbf{y}_{shaped}$ is the shaped one-hot vector. $\mathbf{y}_{shaped}$ is computed as:
\begin{equation}
	\mathbf{y}_{shaped} = Clamp(\mathbf{y}, p_{min}, p_{max})
\end{equation}
\begin{equation}
	p_{min} = (1.0 - p_{max}) / (N_{class} - 1)~~~~~~~~0.5<p_{max}<1.0
\end{equation}
where $Clamp$ is the clamp function, which bounds the minimum and maximum values of a given vector. $\mathbf{y}$ is the classic one-hot label. $p_{max}$ is the parameter that decides the maximum value of $\mathbf{y}$, which is set to 0.6 in the experiment. $N_{class}$ is the number of classes.

\begin{figure}[t]
	\setlength{\abovecaptionskip}{1pt}
 \captionsetup{font={normalsize}}
	\caption{$Mat_p$ and $M_{trojaned}^{\prime}$ computed on trojaned models trained with/without the adaptive attack strategy of posterior shaping. Bright yellow color represents abnormality.}
	\label{fig:adaptive_pos_shaping}
	\captionsetup[subfigure]{justification=centering}
	\captionsetup{font={footnotesize}}
	\centering
	\hspace{-0.2cm}
	\begin{subfigure}{0.24\linewidth}
		\includegraphics[width=\linewidth]{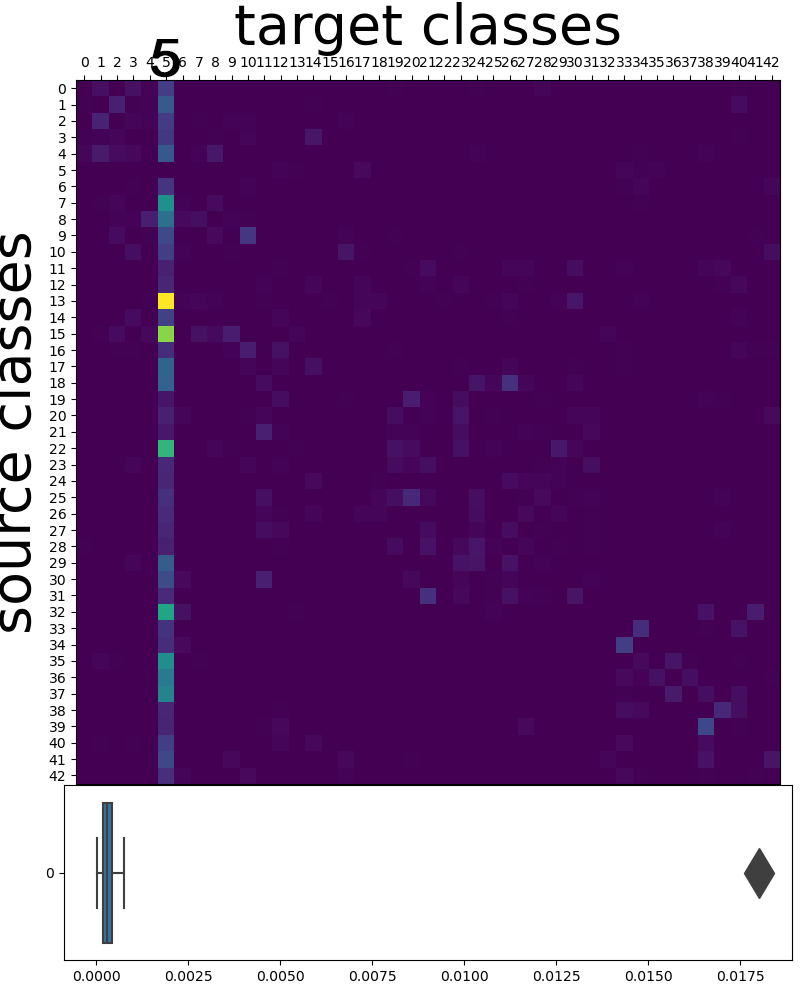}
		\caption{Without Posterior Shaping\\GTSRB\\GoogLeNet\\$M_{trojaned}^{\prime}=(63.958)$}
	\end{subfigure}
	\hspace{-0.2cm}
	\begin{subfigure}{0.24\linewidth}
		\includegraphics[width=\linewidth]{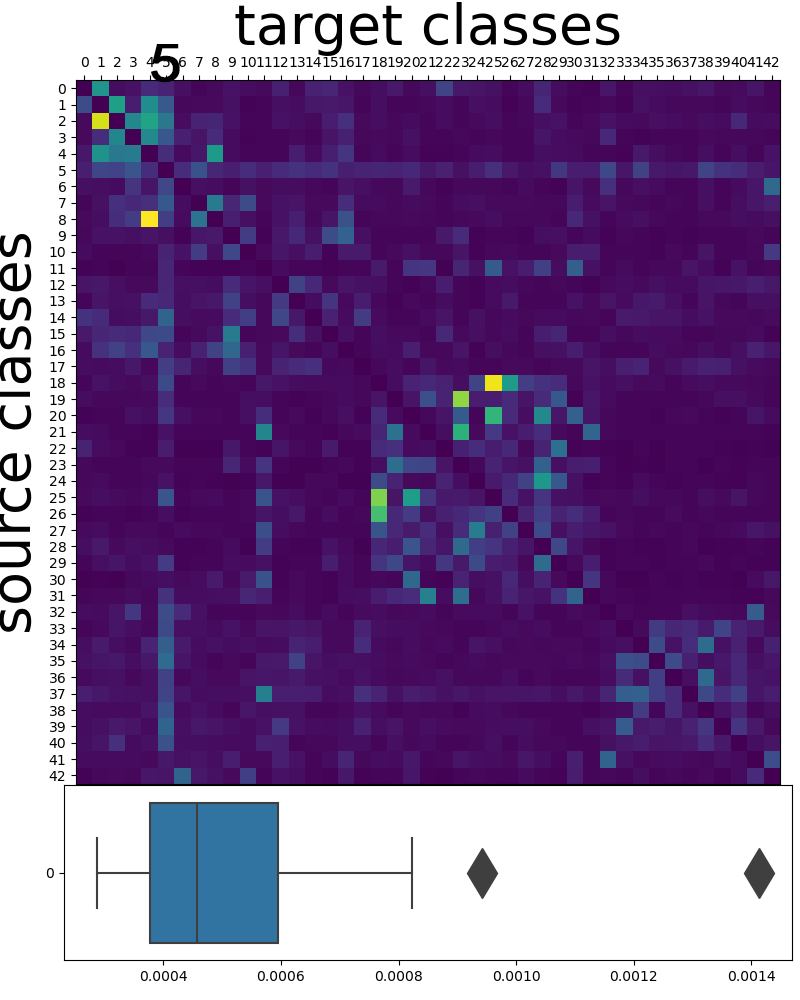}
		\caption{With Posterior Shaping\\GTSRB\\GoogLeNet\\$M_{trojaned}^{\prime}=(2.347)$}
	\end{subfigure}
 \hspace{-0.2cm}
	\begin{subfigure}{0.24\linewidth}
		\includegraphics[width=\linewidth]{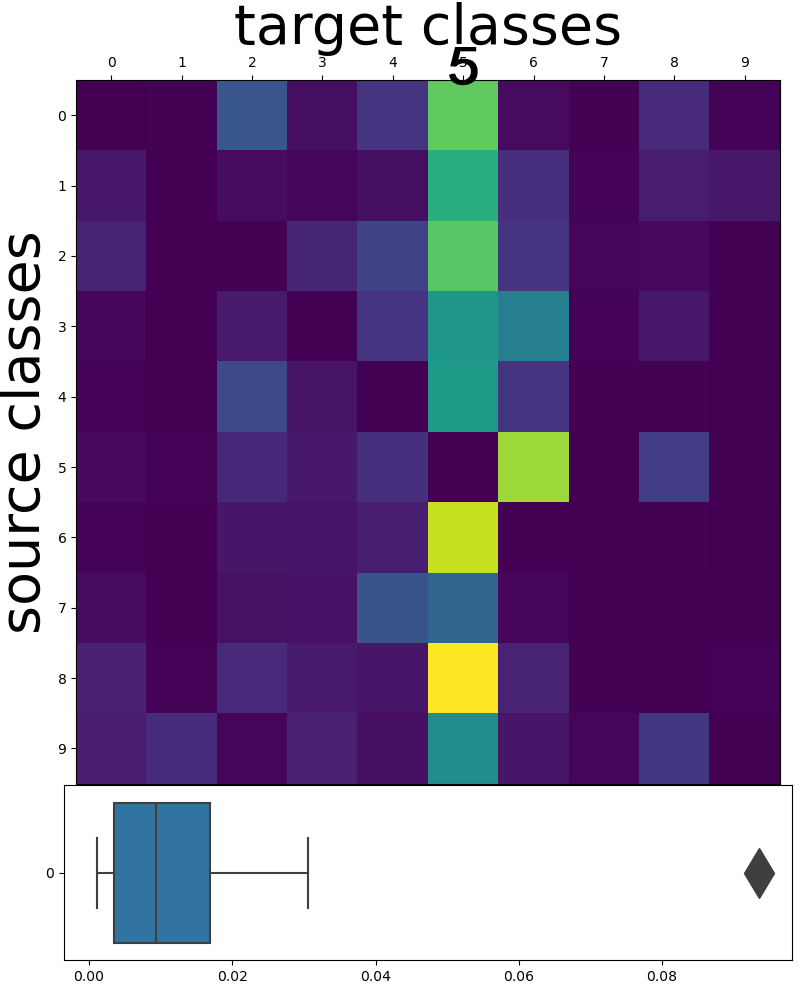}
		\caption{Without Posterior Shaping\\CIFAR-10\\VGG-16\\$M_{trojaned}^{\prime}=(4.080)$}
	\end{subfigure}
 \hspace{-0.2cm}
	\begin{subfigure}{0.24\linewidth}
		\includegraphics[width=\linewidth]{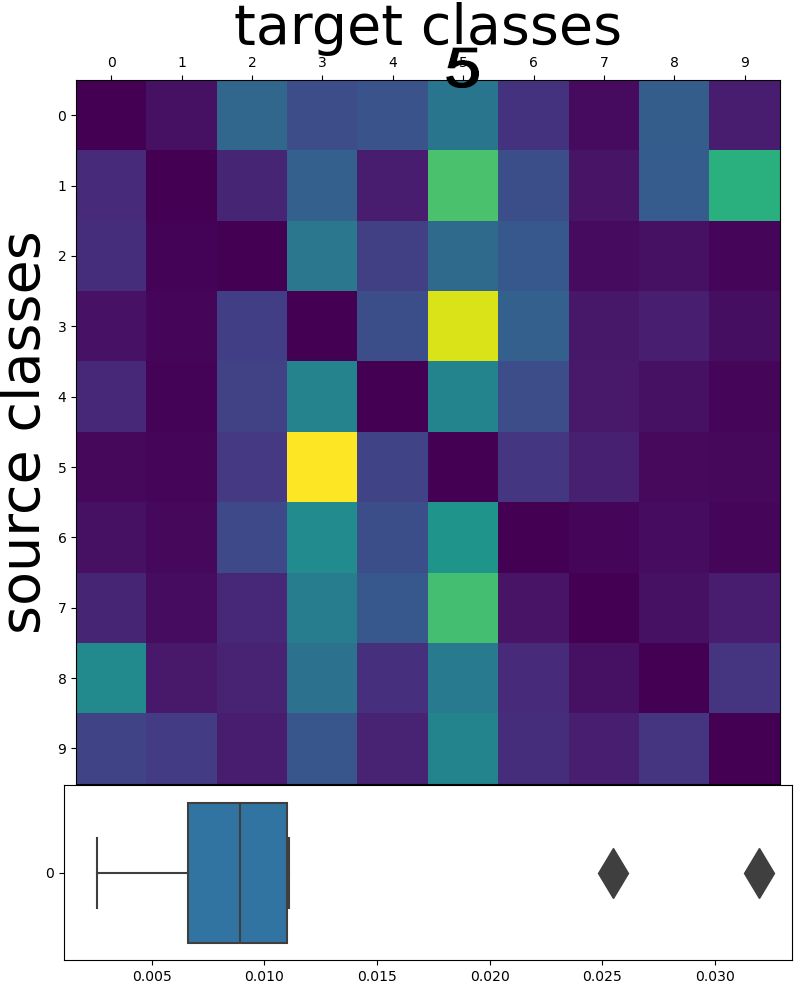}
		\caption{With Posterior Shaping\\CIFAR-10\\VGG-16\\$M_{trojaned}^{\prime}=(3.788)$}
	\end{subfigure}
 \vspace{-0.4cm}
\end{figure}

Some case studies of \system against this adaptive attack are shown in Figure~\ref{fig:adaptive_pos_shaping}. Though posterior shaping does make the trojaned model more evasive against \system, \system manages to outline the target class of the backdoor and computes a relatively large $M_{trojaned}^{\prime}$. Actually, our experiment results show that with this adaptive attack, the TPR/FPR of \system only degrades slightly from 0.99/0.04 to 0.93/0.03 on the GTSRB dataset. On the CIFAR-10 dataset, the TPR/FPR of \system only degrades from 0.88/0.05 to 0.82/0.04. Thus, posterior shaping has minor effect on the defense performance of \system.

 \vspace{-0.3cm}
\subsection{Trojaning the Feature Extractor}\label{sec:adaptive_atk_encoder}
 \vspace{-0.1cm}

\begin{table}[t]	\small
	\centering
 \setlength{\abovecaptionskip}{1pt}
 \caption{\system's defense performance against the adaptive attack strategy of trojaning the feature extractor on the GTSRB and CIFAR-10 datasets. $L_{atk}$ is the layer that the adaptive attacker selects to bypass \system, and $L_{def}$ is the layer selected by \system when inspecting the model.}\label{table:adaptive_atk_encoder}
	\begin{subtable}[t]{0.45\textwidth}
		\centering
		\setlength\tabcolsep{3pt}
       \begin{tabular}{@{}C{2.3cm}C{1.4cm}C{1.4cm}C{1.4cm}@{}}
        \toprule
        \diagbox{$L_{def}$}{TPR/FPR}{$L_{atk}$}  & 10th & 16th & 22nd \\ \midrule
        10th & 0.44/0.02  & 0.79/0.05  & 0.87/0.05  \\ \bottomrule
        \end{tabular}
		\caption{GTSRB, GoogLeNet}\label{table:adaptive_atk_encoder_gtsrb}
	\end{subtable}
	% \quad
 \vspace{-0.1cm}
	\begin{subtable}[t]{0.45\textwidth}
		\centering
        \setlength\tabcolsep{3pt}
		\begin{tabular}{@{}C{2.3cm}C{1.4cm}C{1.4cm}C{1.4cm}@{}}
        \toprule
        \diagbox{$L_{def}$}{TPR/FPR}{$L_{atk}$}  & 8th & 11th & 14th \\ \midrule
        8th & 0.43/0.03  & 0.67/0.04  & 0.76/0.05  \\ \bottomrule
        \end{tabular}
		\caption{CIFAR-10, VGG-16}\label{table:adaptive_atk_encoder_cifar}
	\end{subtable}
  \vspace{-0.2cm}
\end{table}

\begin{table}[t] \small
	\centering
  \setlength{\abovecaptionskip}{1pt}
 \caption{\system's defense performance against non-adaptive attacks when $L_{def}$ is set to different layers.}\label{table:L_non_adaptive_atk}
	\begin{subtable}[t]{0.22\textwidth}
		\centering
		\setlength\tabcolsep{2pt}
       \begin{tabular}{@{}C{0.8cm}C{0.8cm}C{0.8cm}C{0.8cm}@{}}
        \toprule
        $L_{def}$  & 10th & 16th & 22nd \\ \midrule
        TPR & 0.99  & 0.93  & 0.98  \\
        FPR & 0.04  & 0.03  & 0.02  \\
\bottomrule
        \end{tabular}
		\caption{GTSRB, GoogLeNet}\label{table:L_non_adaptive_gtsrb}
	\end{subtable}
	\quad
	\begin{subtable}[t]{0.22\textwidth}
		\centering
        \setlength\tabcolsep{2pt}
		\begin{tabular}{@{}C{0.8cm}C{0.8cm}C{0.8cm}C{0.8cm}@{}}
        \toprule
        $L_{def}$  & 8th & 11th & 14th \\ \midrule
        TPR & 0.88  & 0.83  & 0.82  \\
        FPR & 0.05  & 0.05  & 0.02  \\
\bottomrule
        \end{tabular}
		\caption{CIFAR-10, VGG-16}\label{table:L_non_adaptive_atk_cifar}
	\end{subtable}
  \vspace{-0.2cm}
\end{table}

Inspired by the methodology of BadEncoder~\cite{jia2021badencoder}, the adaptive attacker can trojan the feature extractor part of the model, then fixes the parameters of the feature extractor part and trains the classifier part on the clean training dataset. The detailed adaptive attack method is as follows. (1) The attacker trains a clean model on the clean training dataset, then takes its feature extractor part as $F_{clean}$. (2) The attacker selects clean training samples from the target class as reference samples and records their embeddings, i.e., the outputs of $F_{clean}$. We denote the embedding of a reference sample as $IR_{ref}$. (3) The attacker initializes a feature extractor $F_{bad}$ and trains it with the following two goals. Firstly, for training samples that do not contain the trigger, $F_{bad}$ should output similar embeddings with $F_{clean}$. Secondly, for training samples containing the trigger, the output $\mathbf{y}_{bad}$ of $F_{bad}$ should be similar with a $IR_{ref}$. Specifically, if the training sample contains the trigger, the loss function is defined as the minimum over all candidate $IR_{ref}$, of the average squared difference between $\mathbf{y}_{bad}$ and $IR_{ref}$. (4) After training $F_{bad}$, the attacker initializes a model and replaces its feature extractor part with $F_{bad}$. Finally, the attacker fixes parameters of the feature extractor part then trains the model on the clean training dataset.

By this means, the trojan behavior should be conducted in the feature extractor part rather than the classifier part of the model. Thus, the trojaned model should become more evasive against \system. In the following, we use $L_{atk}$ to denote the layer that the adaptive attacker selects to bypass \system, and $L_{def}$ denotes the layer selected by \system when inspecting the model. As shown in Table~\ref{table:adaptive_atk_encoder_gtsrb}, when $L_{def} = L_{atk}$, the adaptive attack does become more evasive against \system, e.g., the TPR/FPR degrades to 0.44/0.02 when $L_{def}$ and $L_{atk}$ are both 10 on the GTSRB dataset. Meanwhile, if \system inspects a shallower layer than the adaptive attacker, i.e., $L_{def}$ is smaller than $L_{atk}$, \system can still get excellent performance. Thus, to bypass \system, the adaptive attacker should set $L_{atk}$ no larger than $L_{def}$.

\begin{figure}[t]
	\setlength{\abovecaptionskip}{1pt}
 \caption{The trade-off between the evasiveness and the utility of the adaptive backdoor attack of trojaning less layers. ``TPR'' represents the TPR when $L_{def} = L_{atk}$ and FPR is no larger than 0.05. The backdoor setting is ``agnostic backdoor'' and the target class is class 0.}
	\label{fig:tradeoff}
	\captionsetup[subfigure]{justification=centering}
	% \captionsetup{font={footnotesize}}
	\centering
	\begin{subfigure}{0.47\linewidth}
		\includegraphics[width=\textwidth]{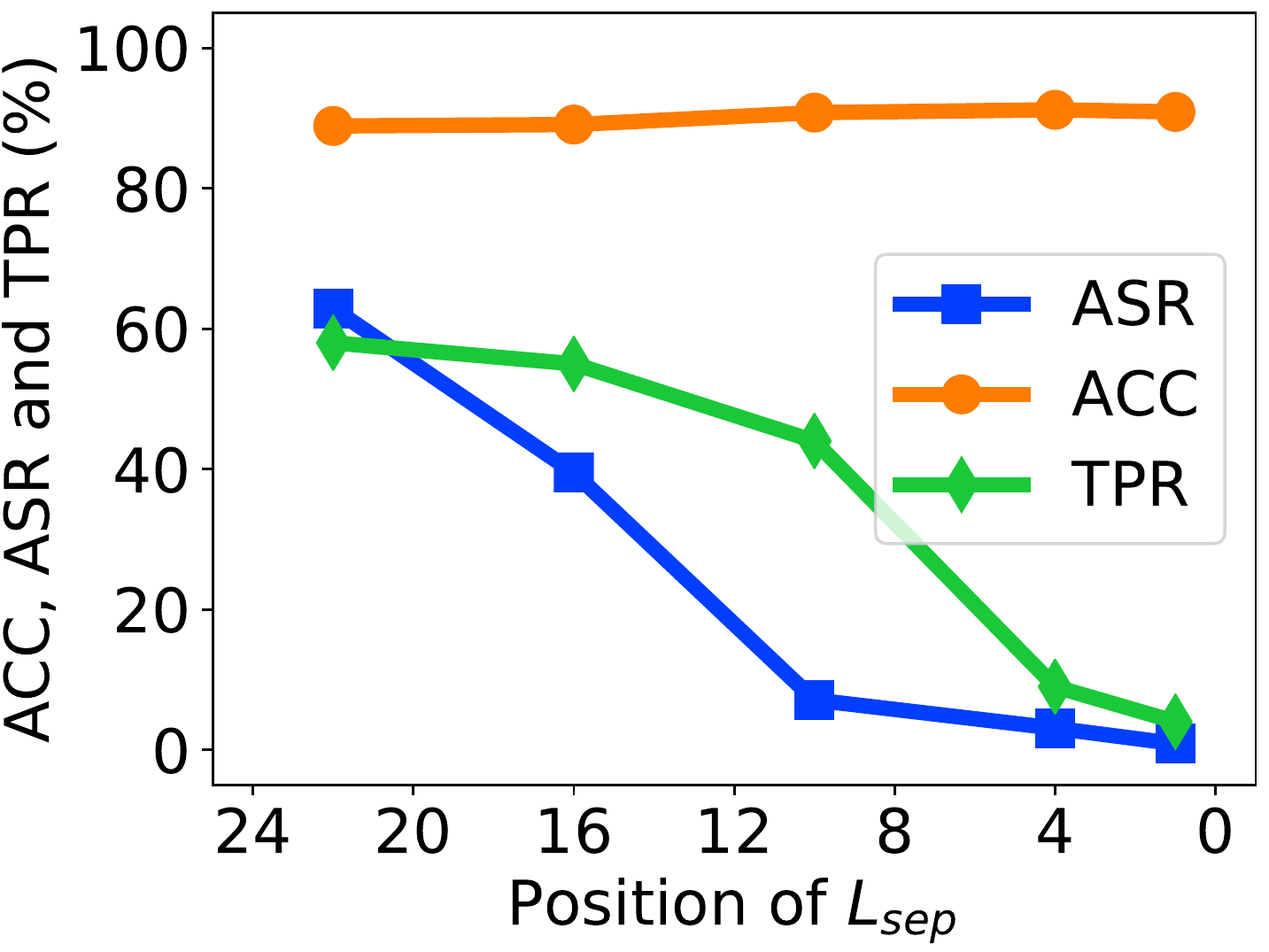}
		\caption{GTSRB, GoogLeNet}\label{fig:tradeoff_gtsrb}
	\end{subfigure}
	\hspace{0.2cm}
	\begin{subfigure}{0.47\linewidth}
		\includegraphics[width=\textwidth]{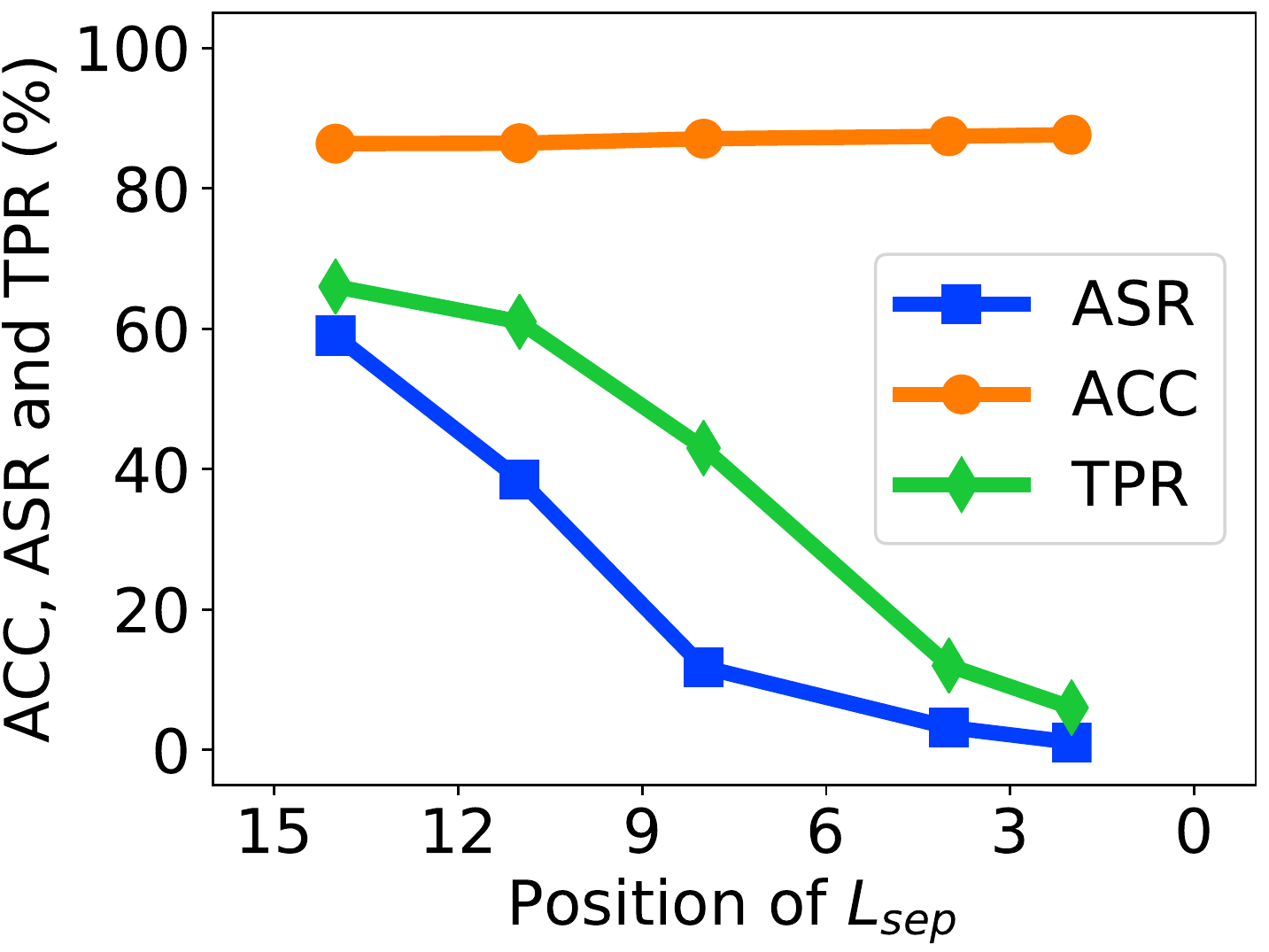}
		\caption{CIFAR-10, VGG-16}\label{fig:tradeoff_cifar}
	\end{subfigure}
	% \captionsetup{font={normalsize}}
  \vspace{-0.3cm}
\end{figure}

However, as shown in Figure~\ref{fig:tradeoff}, as the adaptive attacker sets smaller $L_{atk}$, the attack success rate (ASR) rapidly drops. For example, on the GTSRB dataset, the ASR drops to around 0.05 when $L_{atk}$ is 10. Besides, as shown in Table~\ref{table:L_non_adaptive_gtsrb}, on the GTSRB dataset, when $L_{def}$ is 10, \system can get good defense performance against non-adaptive attacks, i.e., 0.99/0.04 TPR/FPR. To draw a conclusion, to avoid being bypassed by the adaptive attack strategy of trojaning the feature extractor part, \system can inspect the model with $L_{def}$ set to a low value, e.g., the 10th layer of GoogLeNet and the 8th layer of VGG-16. In this way, on the one hand, if $L_{atk}$ is high, the adaptive attacker cannot bypass \system. On the other hand, if $L_{atk}$ is low, the ASR will also be low, i.e., the adaptive attacker will fail to inject a usable backdoor. Similar results can be obtained on the CIFAR-10 dataset, as shown in Table~\ref{table:adaptive_atk_encoder_cifar}, Table~\ref{table:L_non_adaptive_atk_cifar} and Figure~\ref{fig:tradeoff_cifar}.

\vspace{-0.4cm}
\section{Discussion}
\vspace{-0.2cm}
If the trojaned model is trained without data augmentation, \system will become less indicative of discriminating trojaned and benign models, especially for the class-specific backdoors. However, we argue that training trojaned models without data augmentation also reduces the usability of the backdoor. Further, turning off data augmentation is also harmful to the model's performance on the original task.

Another limitation of \system is that its defense performance is limited if the dataset has few classes. The reason is that \system relies on detecting outliers from the posteriors of all classes, and detecting outliers will be more difficult if there are few data points. However, collecting auxiliary labeled data will be much easier if the dataset has few classes, making existing non-data-free backdoor detectors usable. For example, ABS~\cite{liu2019abs} only needs two auxiliary clean labeled samples for binary tasks, one for the negative class and the other for the positive class. Thus, the necessity of data-free methods is reduced for datasets with few classes. Besides, the insights of \system still hold for datasets with few classes, and it is the quantile-related outlier detection step that makes it sensitive to the number of classes. Thus, if we specially design quantile-free anomaly metric for few-class tasks, e.g., using the range of $\mathbf{v}$, i.e., $\max(\mathbf{v})-\min(\mathbf{v})$ as the anomaly metric for two-class tasks, this limitation will be mitigated. We leave the extension of \system to few-class datasets as future work.

\vspace{-0.2cm}
\section{Related Work}
\vspace{-0.1cm}
In addition to the trojan defenses discussed in Section~\ref{sec:existing_trojan_detection}, \system is also related to the following work.

\vspace{-0.2cm}
\subsection{Backdoor Mitigation}
\vspace{-0.1cm}
Different from backdoor detection methods which aim to tell whether a given model is backdoored, backdoor mitigation methods aim to remove the backdoor inside a backdoored model~\cite{liu2018fine, qiu2021deepsweep}. This problem is also named ``backdoor blind removal'' in some literature~\cite{gao2020backdoor}. The backdoor mitigation method is required to remove the backdoor inside a model without a significant drop on the performance of the original task. Neural Cleanse~\cite{wang2019neural} mitigates backdoor via unlearning-based DNN patching, i.e., using inputs with the reversed trigger and correct labels to fine-tune the backdoored model. GangSweep~\cite{zhu2020gangsweep} mitigates backdoor using a similar paradigm with Neural Cleanse, but with the improvement of using a generative adversarial network (GAN) to generate the trigger pattern. AI-Lancet~\cite{zhao2021ai} uses a set of clean-poisoned sample pairs to locate error-inducing neurons in DNN and then fine-tunes or flips the sign of these neurons (weights) to fix the poisoned DNN. Sharing similar insights with AI-Lancet, adversarial neuron pruning (ANP)~\cite{wu2021adversarial} uses a small set of clean labeled samples to locate ``sensitive neurons'' and then prunes these neurons to remove the backdoor. DeepSweep~\cite{qiu2021deepsweep} uses data augmentation techniques to mitigate backdoor attacks as well as enhance the model's robustness.

\subsection{Backdoor-Free Training}
% \vspace{-0.2cm}
Backdoor-free training is an interesting new research topic in neural trojan defenses, aiming to train a clean model even if the training dataset is poisoned. Li et al. proposed Anti-backdoor learning (ABL)~\cite{li2021anti} to train a clean model on a poisoned dataset, with the insight that the DNN model learns backdoored data much faster than learning with clean data.

\section{Conclusion}
In this paper, we present novel insights into the underlying working mechanisms of the trojaned model. We demonstrate that a trojaned model firstly extracts both benign and trigger features, then assign higher priority to the trigger features. Based on the above insights, we propose \system, the first data-free method that is effective against complex neural trojans. Our experiment results on diverse datasets and model architectures show that \system is effective against class-specific/class-agnostic backdoors and various trigger types, even outperforming some non-data-free trojan detectors. We also evaluate \system against possible adaptive attacks and find they cannot effectively bypass \system.

\section{Acknowledgments}
We thank our anonymous reviewers and shepherd for their constructive suggestions. This work was partly supported by the National Key Research and Development Program of China under No. 2022YFB3102100, and NSFC under No.  62102360 and U1936215.

\bibliographystyle{plain}
\bibliography{biblio}

\appendix
\section*{Appendix}
\section{Algorithms of the Vintage-Photography-Style Filter And the Negative Color Filter}
\label{appendix:filter}
\begin{figure}[t]
	\setlength{\belowcaptionskip}{-6pt}
 \captionsetup{font={normalsize}}
	\caption{Comparison of 1-channel images (from the MNIST dataset) with and without the negative color filter.}
	\label{fig:mnist_filter_example}
	\captionsetup[subfigure]{justification=centering}
	\captionsetup{font={footnotesize}}
	\centering
	\begin{subfigure}{0.22\linewidth}
		\includegraphics[width=\linewidth]{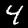}
		\caption{No filter}
	\end{subfigure}
	\hspace{-0.1cm}
	\begin{subfigure}{0.22\linewidth}
		\includegraphics[width=\linewidth]{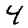}
		\caption{Negative filter}
	\end{subfigure}
	\hspace{-0.1cm}
	\begin{subfigure}{0.22\linewidth}
		\includegraphics[width=\linewidth]{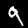}
		\caption{No filter}
	\end{subfigure}
	\hspace{-0.1cm}
	\begin{subfigure}{0.22\linewidth}
		\includegraphics[width=\linewidth]{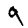}
		\caption{Negative filter}
	\end{subfigure}
\end{figure}

The algorithms of the vintage-photography-style filter and the negative color filter are shown in Algorithm~\ref{alg:filter} and Algorithm~\ref{alg:filter_negative_color}, respectively. Examples of 1-channel images processed by the negative color filter are shown in Figure~\ref{fig:mnist_filter_example}.
%, which is adapted from \url{https://blog.51cto.com/autofelix/4938261}.

\setlength{\textfloatsep}{6pt}
\begin{algorithm}[t]
	\caption{Vintage-photography-style filter}
	\begin{algorithmic}[1]
		\Require{A 3-channel image $img$, parameter $p$ used to adjust how heavy the filter is.}
		\State $img_1 \gets sqrt(img \cdot [1.0, 0.0, 0.0]) \cdot p$
		\State $img_2 \gets img \cdot [0.0, 1.0, 1.0]$
		\State $img \gets img_1 + img_2$ \\
		\Return $img$
	\end{algorithmic}
	\label{alg:filter}
\end{algorithm}

\setlength{\textfloatsep}{6pt}
\begin{algorithm}[t]
	\caption{Negative color filter}
	\begin{algorithmic}[1]
		\Require{A 1-channel image $img$.}
		\State $img \gets 255 \cdot ones\_like(img) - img$ \\
		\Return $img$
	\end{algorithmic}
	\label{alg:filter_negative_color}
\end{algorithm}

\begin{table}[t]
	\setlength{\abovecaptionskip}{6pt}
	\centering
	\caption{Details about the 20 classes in ImageNet-R.}
	\begin{tabular}{@{}ccc@{}}
		\toprule
		\begin{tabular}[c]{@{}c@{}}Class ID \\ in ImageNet-R\end{tabular} &
		\begin{tabular}[c]{@{}c@{}}Class ID in\\ ImageNet\end{tabular}    &
		\begin{tabular}[c]{@{}c@{}}Class\\ Description\end{tabular}                                         \\ \midrule
		0                                                                 & n02114367 & grey wolf           \\
		1                                                                 & n02123159 & tiger cat           \\
		2                                                                 & n02342885 & hamster             \\
		3                                                                 & n02412080 & ram                 \\
		4                                                                 & n02894605 & breakwater          \\
		5                                                                 & n02895154 & breastplate         \\
		6                                                                 & n02930766 & taxicab             \\
		7                                                                 & n02999410 & chain               \\
		8                                                                 & n03089624 & confectionery store \\
		9                                                                 & n03125729 & cradle              \\
		10                                                                & n03141823 & crutch              \\
		11                                                                & n03201208 & dining table        \\
		12                                                                & n03240683 & drilling rig        \\
		13                                                                & n03450230 & gown                \\
		14                                                                & n03773504 & missile             \\
		15                                                                & n03787032 & square academic cap \\
		16                                                                & n03792782 & mountain bike       \\
		17                                                                & n03929855 & pickelhaube         \\
		18                                                                & n03937543 & pill bottle         \\
		19                                                                & n04162706 & seat belt           \\ \bottomrule
	\end{tabular}
	\label{tab:imagenet_r_detail}
\end{table}

\section{Details of Datasets}
\label{appendix:dataset_detail}
The details of the four datasets are listed as follows, including GTSRB, ImageNet-R, CIFAR-10 and MNIST.

\paragraphbe{German Traffic Sign Recognition Benchmark}
German Traffic Sign Recognition Benchmark (GTSRB)~\cite{HoubenGTSRB} is a classic dataset related to the scenario of autonomous driving. The task of GTSRB is to recognize 43 classes of German traffic signs, including the stop sign, several kinds of speed limit signs, and so on. GTSRB is an imbalanced dataset. Specifically, the number of training samples for each class ranges from 210 to 2,250. Thus, we use oversampling to balance the number of samples for each class. After oversampling, each class has 2,250 training samples. For GTSRB, we use GoogLeNet as the model architecture, which is a classic computer vision model proposed in~\cite{szegedy2015going}.

\paragraphbe{ImageNet-R}
ImageNet is derived from the online competition ImageNet Large Scale Visual Recognition Challenge (ILSVRC)~\cite{deng2009imagenet}, which aims to classify samples of 1,000 classes. For simplicity, we randomly choose 20 classes from ImageNet and name this subset of ImageNet as ImageNet-R (restricted). Details of these 20 classes are shown in Table~\ref{tab:imagenet_r_detail}. Each class corresponds to 1,300 training samples and 50 testing samples. We use ResNet-50~\cite{he2016deep} as the model architecture for ImageNet. Samples will be resized to 224$\times$224 before fed to ResNet-50.

\paragraphbe{CIFAR-10}
CIFAR-10~\cite{krizhevsky2009learning} is a classic dataset built for a 10-class image classification task, which has 6,000 samples in the size of 32$\times$32 pixels for each class. Following the common practice of using CIFAR-10, we take 50,000 samples as the training dataset and the other 10,000 as the test dataset. VGG-16~\cite{simonyan2014very} and CNN-7~\cite{lin2020composite} are used as the model architectures for CIFAR-10.

\paragraphbe{MNIST}
MNIST~\cite{lecun1998mnist} is a classic dataset built for handwritten digit recognition, which has 60,000 training samples and 10,000 testing samples. Each sample is a 1-channel image in the size of 28$\times$28 pixels. CNN-7~\cite{lin2020composite} is used as the model architecture for MNIST.

\vspace{0.3cm}
\section{Details About Clean and Trojaned Models Trained for Defense Evaluation}
\label{appendix:details_model_set}
The details of clean and trojaned models trained for defense evaluation are shown in Table~\ref{tab:details_model_sets}. All the clean and trojaned models are trained with standard data augmentations, including randomly cropping and resizing the image, horizontally flipping the image, randomly changing the image's contrast, and random gray scale transformation. Note that the horizontally flipping transformation is disabled for GTSRB because it will change the semantic of some traffic signs. For the ResNet-50 model architecture, we use the ResNet-50 pretrained on ImageNet to accelerate the training process.

\section{Implementation Details of Baseline Methods}
\label{appendix:baseline_details}
\paragraphbe{Code Source}
We use BackdoorBench~\cite{wu2022backdoorbench} to implement baseline methods. The codes can be found at \url{https://github.com/SCLBD/BackdoorBench}. As for ABS, which is not implemented by BackdoorBench, we use its official PyTorch-version code at \url{https://github.com/naiyeleo/ABS/blob/master/TrojAI_competition/round1/abs_pytorch_round1.py}. For DF-TND, we use its official code released at \url{https://github.com/wangren09/TrojanNetDetector}. Note that all baseline methods are implemented with their default parameter settings.

\paragraphbe{Defender's Knowledge}
As an online-detection method, STRIP requires the defender to access inputs with the trigger and a small set of clean samples, so we randomly select 1 image (from the source class) with the trigger and 128 clean images of other classes to inspect a model with STRIP. ABS requires the defender to have a small set of clean samples as seed images, so we randomly select 128 clean images as its seed images. For NC and ANP, we allow them to use the full training dataset and training/testing dataset respectively to maximize their detection ability.

\begin{table*}[t]
	\caption{Details about clean and trojaned models trained to evaluate trojan detection methods. ``Test Acc'' is the model's accuracy of the original task on the clean test dataset. ``ASR'' represents the attack successful rate of the trojan attack. To extensively evaluate \system, we train trojaned models with diverse source/target class settings. For example, on CIFAR-10, for the class-specific backdoor with each trigger type, we train all combinations of source-target class pairs, i.e., at least $9\times10=90$ trojaned models.}
	\begin{tabular}{C{1.5cm}C{1.5cm}C{2.3cm}C{1.6cm}C{1.5cm}C{1.5cm}C{1.5cm}C{1.3cm}C{1.3cm}}
		\hline
		Dataset                     &
		Model                       &
		Trojan Type                 &
		Trigger Type                &
		Source Class                &
		Target Class                &
		Model Quantity              &
		Average Test Acc            &
		Average ASR                                                                                                                                          \\ \hline
		\multirow{7}{*}{GTSRB}      & \multirow{7}{*}{GoogLeNet} & None(Benign)                    &           &      &            & 200 & 90.23\% &         \\ \cline{3-9}
		                            &                            & \multirow{3}{*}{Class-Agnostic} & Patch     &      & 0-42       & 43$\times$4  & 88.96\% & 99.95\% \\
		                            &                            &                                 & Blending  &      & 0-42       & 43$\times$4  & 89.64\% & 99.60\% \\
		                            &                            &                                 & Filter    &      & 0-42       & 43$\times$4  & 88.76\% & 99.83\% \\ \cline{3-9}
		                            &                            & \multirow{3}{*}{Class-Specific} & Patch     & 0-42 & 7,8        & (42$\times$2)$\times$2  & 90.44\% & 99.92\% \\
		                            &                            &                                 & Blending  & 0-42 & 7,8        & (42$\times$2)$\times$2  & 90.08\% & 98.57\% \\
		                            &                            &                                 & Filter    & 0-42 & 7,8        & (42$\times$2)$\times$2  & 88.91\% & 96.93\% \\ \hline
		\multirow{7}{*}{CIFAR-10}   & \multirow{7}{*}{VGG-16}    & None(Benign)                    &           &      &            & 200 & 86.12\% &         \\ \cline{3-9}
		                            &                            & \multirow{3}{*}{Class-Agnostic} & Patch     &      & 0-9        & 10$\times$20  & 84.92\% & 99.86\% \\
		                            &                            &                                 & Blending  &      & 0-9        & 10$\times$20  & 84.95\% & 99.88\% \\
		                            &                            &                                 & Filter    &      & 0-9        & 10$\times$20  & 85.08\% & 98.78\% \\ \cline{3-9}
		                            &                            & \multirow{3}{*}{Class-Specific} & Patch     & 0-9  & 0-9        & (9$\times$10)$\times$2  & 85.69\% & 98.03\% \\
		                            &                            &                                 & Blending  & 0-9  & 0-9        & (9$\times$10)$\times$2  & 86.18\% & 96.42\% \\
		                            &                            &                                 & Filter    & 0-9  & 0-9        & (9$\times$10)$\times$2  & 85.84\% & 95.70\% \\ \hline
		CIFAR-10                    & CNN-7                      & Class-Specific                  & Composite & 0-2  & 0-2        & 3$\times$60  & 83.45\% & 81.24\% \\ \hline
		\multirow{8}{*}{ImageNet-R} & \multirow{8}{*}{ResNet-50} & None(Benign)                    &           &      &            & 200 & 94.74\% &         \\ \cline{3-9}
		                            &                            & \multirow{3}{*}{Class-Agnostic} & Patch     &      & 0-19       & 20$\times$10  & 91.75\% & 99.13\% \\
		                            &                            &                                 & Blending  &      & 0-19       & 20$\times$10  & 92.27\% & 97.83\% \\
		                            &                            &                                 & Filter    &      & 0-19       & 20$\times$10  & 94.02\% & 98.81\% \\ \cline{3-9}
		                            &                            & \multirow{4}{*}{Class-Specific} & Patch     & 0-19 & 0,12,14,18 & (19$\times$4)$\times$2  & 92.06\% & 95.92\% \\
		                            &                            &                                 & Blending  & 0-19 & 0,12,14,18 & (19$\times$4)$\times$2  & 94.43\% & 99.87\% \\
		                            &                            &                                 & Filter    & 0-19 & 0,12,14,18 & (19$\times$4)$\times$2  & 93.20\% & 97.96\% \\
		                            &                            &                                 & Natural   & 13   & 0          & 200  & 92.72\% & 91.34\% \\ \hline
		\multirow{7}{*}{MNIST}      & \multirow{7}{*}{CNN-7}     & None(Benign)                    &           &      &            & 200 & 98.65\% &         \\ \cline{3-9}
		                            &                            & \multirow{3}{*}{Class-Agnostic} & Patch     &      & 0-9        & 10$\times$20  & 96.94\% & 99.69\% \\
		                            &                            &                                 & Blending  &      & 0-9        & 10$\times$20  & 96.92\% & 99.82\% \\
		                            &                            &                                 & Filter    &      & 0-9        & 10$\times$20  & 97.43\% & 99.98\% \\ \cline{3-9}
		                            &                            & \multirow{3}{*}{Class-Specific} & Patch     & 0-9  & 0-9        & (9$\times$10)$\times$2  & 97.52\% & 99.21\% \\
		                            &                            &                                 & Blending  & 0-9  & 0-9        & (9$\times$10)$\times$2  & 97.73\% & 99.38\% \\
		                            &                            &                                 & Filter    & 0-9  & 0-9        & (9$\times$10)$\times$2  & 97.61\% & 99.38\% \\ \hline
	\end{tabular}
	\label{tab:details_model_sets}
\end{table*}

\end{document}